# Characterization of Exoplanet Atmospheres with the Optical Coronagraph on *WFIRST*

B. Lacy[1] , D. Shlivko[2], and A. Burrows[1]

[1] Department of Astrophysics, Princeton, NJ 08540, USA
[2] California Institute of Technology, Pasadena, CA 91125, USA; blacy@princeton.edu

## Abstract

*Wide-Field Infrared Survey Telescope* (*WFIRST*)-CGI is a NASA technology demonstration mission that is charged with demonstrating key technologies for future exo-Earth imaging missions in space. In the process, it will obtain images and low-resolution spectra of a handful to a dozen extrasolar planets and possibly protoplanetary disks. Its unprecedented contrast levels in the optical will provide astronomers' with their first direct look at mature, Jupiter-sized planets at moderate separations. This paper addresses the question: what science can be done with such data? An analytic noise model, which is informed by the ongoing engineering developments, is used to compute maximum achievable signal-to-noise ratios and scientifically viable integration times for hypothetical star–planet systems, as well as to investigate the constraining power of various combinations of *WFIRST*-CGI photometric and spectral observations. This work introduces two simple models for planetary geometric albedos, which are inspired largely by the solar system's gas giants. The first planet model is a hybrid Jupiter–Neptune model, which separately treats the short and long wavelengths where chromophores and methane dominate absorption, respectively. The second planet model fixes cloud and haze properties in *CoolTLusty* to match Jupiter's albedo spectrum, it then perturbs only the metallicity. MCMC retrievals performed on simulated observations are used to assess the precision with which planet model parameters can be measured subject to different exposure times and observing cases. Planet radius is recovered within ±15% for all observing cases with both the hybrid model and the *CoolTLusty* metallicity grid. Fit results for both models' parameterizations of geometric albedo spectra demonstrate that a rough indication of the metallicity or methane content should be possible for some *WFIRST*-CGI targets. We conclude that real observations will likely be able to differentiate between extreme cases using these models, but will lack the precision necessary to uncover subtle trends.

## 1. Introduction

Space-based high-contrast imaging promises to inaugurate a new phase of exoplanet atmospheric studies. Initial efforts from the ground have successfully imaged young self-luminous, wide-separation exoplanets (and brown dwarfs) in the infrared, where planet–star contrast ratios are on the order of $10^{-4}$ to $10^{-6}$, (GPI: Macintosh et al. 2006, 2015; SPHERE: Mesa et al. 2015). NASA's *Wide-Field Infrared Survey Telescope* (*WFIRST*) mission is likely to fly an optical coronagraph (CGI) that is capable of detecting and characterizing cool giant exoplanets (EGP's) around nearby stars at separations of ∼1 to ∼6 au, where optical contrast ratios are on the order of $10^{-8}$ to $10^{-10}$ (Spergel et al. 2013). *WFIRST*-CGI is defined as a technology demonstration mission, which means that simply attaining the specified contrasts (currently around $5 \times 10^{-8}$, Douglas et al. 2018) and other instrument capabilities constitutes mission success. Nonetheless, this is an exciting opportunity to obtain images and low-resolution optical albedo spectra at unprecedented contrast ratios. The engineering team's current best estimates put the contrast an order of magnitude lower than mission requirements, around $10^{-9}$ (Mennesson et al. 2018). This advance in observational capability raises the theoretical question: what new insights into giant planets can be gleaned from such data? This optical range can be a powerful tool for probing giant exoplanet characteristics. It spans prominent methane features near ∼0.62, ∼0.74, ∼0.81, and ∼0.89 μm, ammonia features at ∼0.65 and ∼0.79 μm, and a broad water band at ∼0.94 μm. In addition, Rayleigh scattering off molecules and Mie-like scattering off cloud particulates and hazes can modify the planet's reflectivity in diagnostic ways (Burrows et al. 2004; Sudarsky et al. 2005). As the design process converges, it is timely to consider the degree to which *WFIRST*-CGI low-resolution spectra and imaging will be able to probe these anticipated spectral signatures.

A number of studies have already begun to explore the expected capabilities of *WFIRST*-CGI. These studies build on well-established reflection spectra theory (Marley et al. 1999; Sudarsky et al. 2000), the literature surrounding the viability of space-based coronagraph or external occulter surveys (Agol 2007; Brown & Soummer 2010; Savransky et al. 2010; Turnbull et al. 2012; Stark et al. 2014), and the ongoing engineering efforts to design and build space-based external occulters and coronagraphs, in general, and for *WFIRST*-CGI, specifically (Kasdin et al. 2003; Soummer 2005; Carlotti et al. 2011; Zimmerman et al. 2015; Balasubramanian et al. 2016; Cady et al. 2016; Krist et al. 2016). Much work has gone into estimating the yield of the imaging portion of the mission. Greco & Burrows (2015) performed Monte Carlo simulations to quantify the detectability of giant planets under various conditions in the context of *WFIRST*, (see also Savransky & Garrett 2016; Traub et al. 2016; Garrett et al. 2017). Many quantitative predictions of yields in these studies depend upon assumptions of instrument performance and extrapolations of underlying exoplanet populations into regimes for which we do not yet have sufficient data. Explorations of the integral field spectrograph (IFS) capabilities have lagged behind imaging because they require additional assumptions about likely planet atmospheres. Cahoy et al. (2010) studied how optical reflection spectra and colors vary with planet/star separation, metallicity, mass, and observed phase. Lupu et al. (2016) developed an atmospheric retrieval methodology for optical reflection spectra and demonstrated its abilities on simulated direct imaging data. Their geometric albedo models allowed for one or two cloud layers, variations in surface gravity, and variations in methane abundance. Robinson et al. (2016) presented a general model for any space-based coronagraph equipped with an IFS and computed the required integration times to reach a given signal-to-noise ratio (S/N) as a function of host star effective temperature and wavelength for a variety of planet characteristics. Nayak et al. (2017) used the noise model of Robinson et al. (2016) to simulate *WFIRST*-CGI data and extended the work of Lupu et al. (2016) to consider how uncertainties in planet radius and phase angle influence their ability to fit similar models.

As the design process has continued, the assumptions made in Robinson et al. (2016) and Nayak et al. (2017) can now be improved. Nemati et al. (2017) present an analytic parameterized noise model that is tailored particularly to *WFIRST*-CGI observations, utilizing more recent instrument specifications and exploring how mission performance depends on these instrument specifications. Their particular focus was on raw contrast, post-processing, and core throughput. Rizzo et al. (2017) conducted a rigorous simulation of the IFS + EMCCD performance. Their publicly available software, `crispy`, takes a high-resolution astrophysical-observing-scenario-dependent data cube as input and then outputs a data cube that is consistent with the *WFIRST*-CGI data products. These resulting data cubes can then be used to test post-processing methods, observing scenarios, and instrumental parameters in an effort to optimize the science yield of the mission.

In this paper, we create a versatile pipeline to simulate *WFIRST*-CGI observations, which can be adapted to explore new planet models and target system geometries, and can be updated with the latest expected instrument capabilities as they are finalized. The low throughput of the *WFIRST*-CGI system and the low spectral resolution that this necessitates for the IFS will yield data that are unlikely to constrain detailed atmosphere models. We thus present two appropriately simple planet models as input for our pipeline: a hybrid Jupiter–Neptune albedo spectrum, and a suite of reflection spectra generated with *CoolTlusty* (Hubeny & Lanz 1995; Sudarsky et al. 2003; Burrows et al. 2006) which fix cloud parameters tuned to match Jupiter at a metallicity of 0.5 dex (around $3.16 \times$ solar), and then perturb the metallicity between 1 and 30 times solar values. Both classes of geometric albedo model are paired with Jupiter's measured phase function (Mayorga et al. 2016). Rather than utilizing the detailed model of Rizzo et al. (2017) in our pipeline, we chose to adopt the noise model of Nemati et al. (2017) to simulate observed spectra. We then carry out MCMC parameter retrievals to assess the recovery of underlying planet model parameters. This allows us to quickly explore many variations of our planet models, many planet–star scenarios, and compare results for several combinations of *WFIRST*-CGI's imaging and spectral coverage. We explore both the originally proposed imaging and spectral bands, and those that have been selected for full commissioning in the scaled back version of the mission that is now adopted.

In Section 2, we provide a brief review of the portions of the reflection spectra theory that are most relevant to simulating and interpreting direct images of exoplanets. Section 3 outlines the *WFIRST*-CGI noise model that is implemented in our pipeline, while the detailed calculations are given in the Appendix. Section 4 describes the two planet models that are used as inputs for our pipeline. Section 5 presents calculations of integration times to achieve different S/N's and parameter estimate precisions for varying S/N's, integration times, and observing cases. Section 6 discusses the implications of these results for giant exoplanet science in the era of *WFIRST*-CGI. Finally, Section 7 summarizes our work, highlighting our important conclusions.

## 2. Relevant Reflection Spectra Theory

To investigate the relationship between a planet's atmospheric properties and the resulting observations from a space-based telescope such as *WFIRST*, one must focus primarily on modeling the geometric albedo spectrum and deriving the resulting planet–star flux ratios. For mature EGP's whose thermal emissions in the optical are negligible, the emergent flux from the planet comes almost entirely from reflection of stellar light. Thus, the wavelength-dependent planet–star flux ratio is merely the fraction of stellar light reflected toward the observer by the EGP. This depends trivially on the planet's size and orbital distance, but is also a more intricate function of the planet's phase and atmospheric composition.

In addition to these orbital and physical determinants of the flux ratio spectrum, some systems may present themselves with a number of more exotic and subtle factors that are not presently reflected in our models. For instance, additional dependencies on orbital inclination may arise from the presence of rings, which for Saturn can vary its luminosity by up to a factor of two (Mallama 2012), or from rotation-induced oblateness (Mallama et al. 2017), which influences brightness according to the area oriented to reflect light toward Earth. Latitudinal variations in atmospheric composition (Schmude et al. 2015; Mallama et al. 2017) may add further to this inclination dependence. Other corrections may be needed to account for intrinsic temporal brightness variations due to rotation and/or evolution of atmospheric structure, as is seen for Neptune (Schmude et al. 2016). Lastly, though planetary phase functions for the solar system planets are not well-known, we do know that opposition effects, glories, and rainbows are possible for a range of atmospheric and surface characteristics. We do not attempt to model or incorporate any of these possibilities, concluding that such detailed models are premature at this stage in exoplanet reflection science.

### 2.1. Albedos and Phase Functions

Formally, the proportion of incident light of wavelength $\lambda$ reflected by a planet at a phase angle $\alpha$ toward an observer is governed by the planet's geometric albedo $A_g(\lambda)$ and its phase function $\Phi(\alpha, \lambda)$, where $\alpha$ is the exoplanet-centric angle between the star and the observer (Sudarsky et al. 2000; Madhusudhan & Burrows 2012; Greco & Burrows 2015). The geometric albedo measures the reflectivity of the planet at full phase ($\alpha = 0$), while the phase function is normalized to 1 at full phase and dictates the variation of reflectivity with phase. In terms of these quantities, the planet–star flux ratio is given by

$$\frac{F_p}{F_\star}(\alpha, \lambda) = \frac{R_p^2}{a^2} \cdot A_g(\lambda) \cdot \Phi(\alpha, \lambda), \tag{1}$$

where $R_p$ is the planet's radius and $a$ is its orbital distance.

A planet's geometric albedo and the behavior of its phase function are determined by the scattering properties of its atmosphere, which in turn depend on the atmosphere's composition. Cloud layers in the atmosphere provide the primary source of scattering, giving the planet a non-negligible albedo (typically on the order of a few tenths). Scattering effects from hazes and gases further increase the albedo, but their influence is less substantial. Clouds and hazes also provide sources of absorption in the atmosphere, although the effect is modest relative to that of gases. The absorption of gases is also far more wavelength-dependent than that of other atmospheric components, which makes gases the primary sculptors of the shape of the geometric albedo spectrum (Sudarsky et al. 2000, 2003, 2005).

The subtle dependence of a planet's phase function $\Phi(\alpha, \lambda)$ on wavelength has important consequences for spectral observations of an exoplanet near quadrature ($\alpha = 90°$). This phenomenon is best illustrated by the phase dependence of a planet's color indices (Sudarsky et al. 2005; Cahoy et al. 2010; Mayorga et al. 2016), which may vary by up to one magnitude (through Johnson–Cousins filters) between new and full phases for an EGP in an edge-on orbit.

### 2.2. Time-dependent Observables

It is possible to substitute the dependencies of the planet–star flux ratio on orbital distance and phase with more standardized orbital parameters, including inclination, eccentricity, argument of periastron, and time elapsed since periastron. This substitution requires a mapping that gives the planet's phase angle $\alpha$ as a function of time. We follow the formalism summarized in Madhusudhan & Burrows (2012), which draws on many previous works, including Russell (1916), Horak (1950), van de Hulst (1974), Seager & Sasselov (1998), Marley et al. (1999), Sudarsky et al. (2000), Seager et al. (2000), Stam et al. (2004), and Cahoy et al. (2010). We first calculate the planet's mean anomaly $M$ from its definition (see Murray & Dermott 1999, or other textbooks covering orbital dynamics):

$$M = 2\pi \cdot \frac{t}{P}, \tag{2}$$

where $t$ is the time since periastron and $P$ is the period. Next, to determine the exoplanet's eccentric anomaly, $E$, we utilize Kepler's equation:

$$M = E - e \cdot \sin(E), \tag{3}$$

where $e$ is the orbit's eccentricity. Note that because $M$ increases monotonically with $E$ ($\frac{dM}{dE} = 1 + e \cdot \cos(E)$ and $0 \leqslant e < 1$), there will always exist a unique solution for $E$. The true anomaly $\theta$ is then related to the eccentric anomaly via

$$\theta = 2 \cdot \tan^{-1}\left(\sqrt{\frac{1+e}{1-e}} \cdot \tan(E/2)\right). \tag{4}$$

Finally, the phase angle can be retrieved using

$$\alpha = \cos^{-1}(\sin(i) \cdot \sin(\theta + \omega_p)), \tag{5}$$

where $i$ is the orbit's inclination (0° representing a face-on orbit) and $\omega_p$ is the argument of periastron.

Using this formalism, observables such as exoplanetary colors and planet–star flux ratios can be generated as a function of time for an orbit with known Keplerian elements. This allows us to simulate observations using measured properties of known radial velocity (RV) planets around nearby stars, which are the most likely targets for *WFIRST*-CGI (Traub et al. 2016).

### 3. *WFIRST*-CGI Mission and Noise Model

*WFIRST*-CGI is a NASA technology demonstrator that aims to pave the way for future exo-Earth imaging missions. It will demonstrate many key technologies for the first time in space, including high actuator count deformable mirrors, low-noise, single photon-counting detectors in the visible, new coronagraph masks and architectures, low-resolution integral field spectroscopy, advanced algorithms for wavefront sensing and control, high-fidelity integrated spacecraft and coronagraph modeling, and post-processing approaches to extract images and spectra (Mennesson et al. 2018). Before it can be deemed to be a success, *WFIRST*-CGI must first meet a number of requirements that are outlined in Douglas et al. (2018). Two of these requirements are: (1) the instrument should be able to measure the brightness of a point source to an S/N of 10 or greater within 10 hr of integration time, and (2) the instrument should be able to measure the spectra of said source with $R = 50$ or greater spectral resolution to an S/N of 10 within 100 hr of integration time on target. This must be possible for objects with source-to-star flux ratios as faint as $5 \times 10^{-8}$–$1 \times 10^{-7}$, and for apparent separations from 0.21 to 0.6 arcsec. In the event of the successful completion of its technology demonstration, NASA is committed to making *WFIRST*-CGI available to the general astronomy community as a science instrument. If the capabilities progress as hoped, then observers will be able to obtain visible light spectroscopy and photometry of known young, self-luminous planets, proto-planetary and debris disks, and some known RV planets.[3] The mission is sensitive to cool, moderate-separation EGPs, which may allow it to probe the structure and composition of planetary systems unlike those best studied by transit and ground-based IR direct imaging.

To accomplish all of these tasks, *WFIRST*-CGI will be equipped with two coronagraphs: a hybrid lyot coronagraph (HLC, Trauger et al. 2016), and a shaped-pupil coronagraph (SPC, Balasubramanian et al. 2016). The HLC will be paired with imaging capabilities and will have an annular field of view extending to small inner working angles. This makes it well-suited to blind planet searches. The SPC will be able to operate in two modes. One mode creates a "bowtie" shaped field of view at smaller working angles, suitable for characterizing known planets when paired with the IFS. The second mode creates an annular field of view at larger working angles, which is suitable for characterizing extended disks when paired with the imager. Engineers are optimistic that both the HLC and SPC coronagraphs will be able to achieve contrast ratios on the order of $10^{-9}$, although the SPC design with an annular field of view will have a brighter floor. Figure 1 shows a model of the two coronagraphs' performances as a function of working angle. These are the current best estimates, which we use throughout our analysis. We emphasize, as stated in the previous paragraph, the performance required to deem the technology demonstration a success is at roughly 1.5 orders of magnitude worse than these best estimates (Douglas et al. 2018).

---

[3] *WFIRST* white paper by Vanessa Bailey, available at: https://caltech.app.box.com/v/nas-cgi.

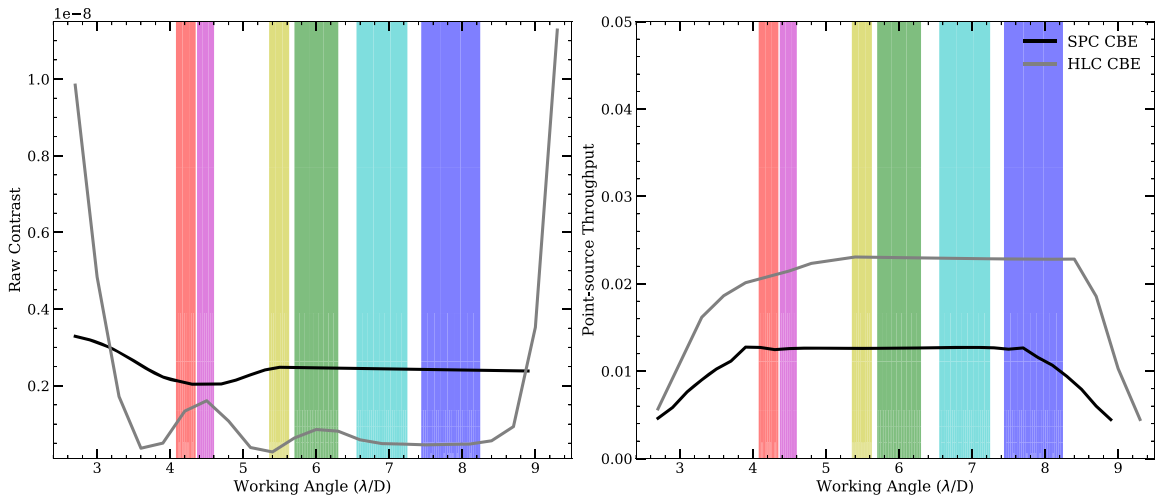

**Figure 1.** Azimuthal mean raw contrast and point-source throughput of the SPC (black line) and HLC (gray line) as a function of working angle in units of $\lambda/D$. Models are generated with the numerical method described in Krist et al. (2016), but come from more recent tests than those shown in the publication. The point-source throughput shown here describes the fraction of a target's incident light that ends up in the core of the PSF (the region with values >50% of the PSF peak value). It includes losses from pupil obscurations, focal plane masks, reflections, and filters, but not losses from polarizers, detectors etc. The SPC design shown here is the "bow-tie" shaped field of view which will be used with the IFS. The HLC design will be used for imaging at wavelengths below 0.721 $\mu$m. A variation of the SPC with an annular field of view will be used for the longer wavelength imaging bands. In practice, the imaging will likely be done with a polarizer, which reduces the core throughput of the HLC to around the level of the SPC and allows diagnostic measurements of the polarization of planet light. These curves can be read as wavelength-dependent performance for a given planet–star system if one converts the fixed working angle of the planet in radians into appropriate $\lambda/D$ units for the wavelength of interest. The imaging bands are shaded and labeled with their central values at the working angles that they would have for our fiducial system: a planet on a circular orbit, located 3.8 au from its primary, observed from a distance of 10 pc, at $\alpha = 65°$, giving it a working angle of 0.38 arcsec. This fiducial system was chosen to fall within the working angles of both coronagraph models across the full range of wavelength coverage.

Before de-scoping due to budgetary constraints, the HLC and SPC (with the annular field of view) were intended to be used with six imaging bands spanning roughly 0.5–1.0 $\mu$m with spectral widths ranging from 5% to 10%. The SPC (with the "bow-tie" field of view) was also intended to be paired with an IFS, which would have been capable of dispersing light with resolving power of $R \sim 50$ in three slightly overlapping bands spanning 0.6–1.0 $\mu$m with $\sim$18% width (Mandell et al. 2017; Saxena et al. 2017). Figure 2 shows these, initially planned, imaging and IFS band locations and widths.

At the time of writing, the HLC is intended for use with a 10% width imaging band at 0.575 $\mu$m, the SPC (with an annular field of view) is intended for use with a 10% width imaging band centered at 0.825 $\mu$m, and the SPC (with the "bow-tie") field of view is intended for use with two $\sim$18% width IFS bands centered at 0.66 and 0.76 $\mu$m. For the technology demonstration mission requirements, only the IFS band centered at 0.76 $\mu$m, the imaging band centered at 0.575 $\mu$m, and the imaging band centered at 0.825 $\mu$m will undergo the thorough testing that is required to be fully commissioned. A subset of the other filters may still be manufactured and may fly with the mission, but there is less certainty as to their ultimate performance.[4]

In this work, we choose to explore the science possible with both the originally planned *WFIRST*-CGI spectral coverage and

---

with an approximation of the de-scoped mission capabilities. We do not model the 0.825 $\mu$m band precisely because we have not yet incorporated the SPC with the annular field of view format into our model.

Due to the low flux levels, extremely low throughput, and long integration times under which *WFIRST*-CGI will operate, the e2v CCD201-20 EMCCD has been selected as the detector for both the CGI imaging camera and the integral field spectrograph (Harding et al. 2016). The EMCCD can be operated in a photon-counting mode, which essentially nullifies read noise (Denvir & Conroy 2003; Nemati 2014; Harding et al. 2016) and allows the detection of single photons. A wide variety of observing strategies and post-processing techniques have been developed for the direct imaging of exoplanets. The exact procedure adopted by *WFIRST*-CGI will depend on what advances are made in the upcoming years. For our simulated observations, we assume that observations are carried out to allow a final data analysis with reference differential imaging (RDI, Ygouf et al. 2015). With this in mind, a typical observing procedure for a spectroscopic or imaging target could follow four steps (Krist et al. 2016; Nemati et al. 2017):

1. Observe a nearby star of the same brightness and approximate color as the target to obtain a reference PSF.
2. Observe a nearby star that is significantly brighter than the target and calibrate the wavefront control system to optimize starlight suppression in the target region (this procedure is commonly referred to as "digging the dark hole" in direct imaging literature).
3. Observe the target.
4. Return to the nearby star of the same brightness and color to carry out additional observations, which will then be used to remove speckles via post-processing.

Because speckles can mimic planets, the thermal stability of the instrument places constraints on the total time that an

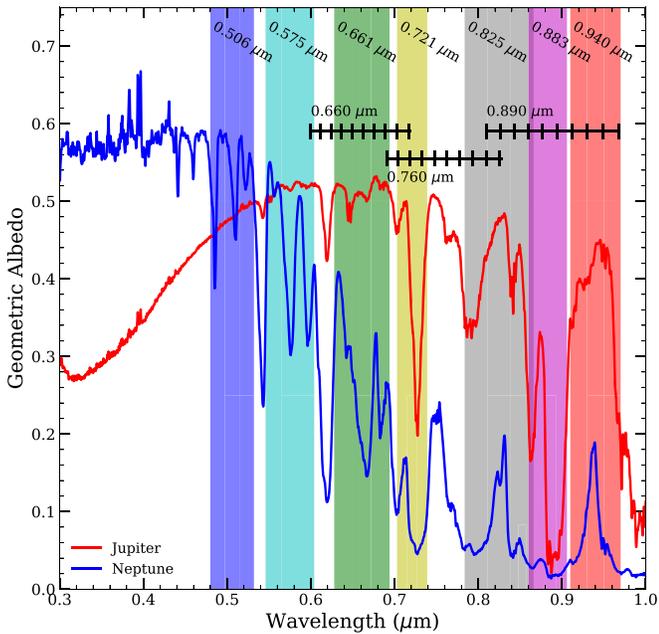

**Figure 2.** The wavelengths of the formerly planned and currently planned *WFIRST*-CGI bandpasses. Three overlapping IFS bands are shown with black horizontal lines: a 18.2% full-width band centered at 0.660 μm, a 18.2% width band centered at 0.760 μm, and a 18.0% width band centered at 0.890 μm. Vertical hash marks delineate a resolution of $R = 50$. Imaging filters are shown as shaded bands and labeled with their central wavelengths: 0.506, 0.575, 0.661, 0.825, 0.883, 0.721, and 0.940 μm, with corresponding widths of: 10.3%, 10.1%, 10.0%, 5.2%, 10.0%, 5.0%, and 6.4%, respectively. We emphasize that the 0.76 μm IFS bandpass, 0.575 μm imaging bandpass, and 0.825 μm imaging bandpass are the *only* filters that will undergo full commissioning. The IFS filter centered at 0.66 μm will be included but it will not be tested. At the time of writing, the other originally planned filters are not set to be included in the mission. For the purposes of this work, all bands are assumed perfect top-hat functions between 0% transmission and 90% transmission. *WFIRST*-CGI coverage is overlaid by Karkoschka's (1994) measurements of Jupiter's and Neptune's geometric albedo spectra to provide a sense of the spectral features that each band may be sensitive to.

**Table 1**
Summary of Adopted Fiducial Model Parameters, Including Instrument Parameters and Astrophysical Parameters

| Parameter | Fiducial Value |
|---|---|
| $f_{pp}$ | 1/12 |
| $A_{PM}$ | 3.684 (m²) |
| $D$ | 2.37 (m) |
| $\tau_{obs}$ | 0.835 |
| $\tau_{ref}$ | SPC: 0.383, HLC: 0.573 |
| $\tau_{BI}$ | 0.90 |
| $\tau_{pol}$ | 1.0 (would be ∼0.5 if a polarizer were used) |
| $\tau_{occ}$ | SPC: 0.13, HLC: 0.42 (radial average values) |
| $\tau_{core}$ | SPC: 0.037, HLC: 0.043 (radial average values) |
| $I_{pk}$ | SPC: 0.0013, HLC: 0.0050 (radial average values) |
| $C_{raw}$ | see Figure 1 left-hand panel |
| $\Theta_{IWA}$ | ∼3.0 ($\lambda/D$) |
| $\Theta_{OWA}$ | ∼9.0 ($\lambda/D$) |
| $\eta$ | ∼0.65, see Section A.1 |
| $R$ | 50.0 |
| $N_{lens}$ | 3.0 |
| $N_{spec}$ | 5.0 |
| $N_{pix}^{detector}$ | imaging: 11, IFS: 45 |
| $\lambda_c$ | imaging: 0.508 μm, IFS: 0.66 μm |
| $i_d$ | $4.6 \times 10^{-4}$ ($e^-$/pixel/s) |
| $i_r$ | $1.7 \times 10.0^{-6}$ ($e^-$/pixel/frame) |
| $q_{cic}$ | 0.016 ($e^-$/pixel/frame) |
| $t_{fr}$ | imaging: 10 (s), IFS: 80 (s) |
| $F_{0,V}$ | $3.6 \times 10^{-8}$ (W m⁻² μm⁻¹) |
| $F_{\odot,V}$ (1 au) | $1.86 \times 10^3$ (W m⁻²) |
| $M_{z,V}$ | 23.0 (magnitudes arcsec⁻²) |
| $M_{ez,V}$ | 22.0 (magnitudes arcsec⁻²) |
| $F_{\odot,\lambda}$ (1 au) | from Kurucz (2005)[a] |

**Notes.** We assume the current best estimates of the engineering team as of 2018 July, and the detector performance corresponding to damage acquired from three months of operation at L2.
[a] http://kurucz.harvard.edu/stars/Sun/fSunallp.1000resam251

observation of a single object can last (including time on target and time calibrating). Exposure times exceeding 500 hr for a single IFS bandpass and 100 hr for a single imaging bandpass are thus undesirable (*WFIRST*-CGI engineering team 2018, private communication).

Our noise model closely follows the work of Nemati et al. (2017), with some slight adaptations. Here, we provide a brief overview of that model. Detailed calculations of contributing noise terms, throughputs, and quantum efficiencies are included in the Appendix. Our model represents a space-based telescope with a primary mirror collecting area $A_{PM}$ that is equipped with a coronagraph achieving raw contrast $C_{raw}$, planetary, zodiacal, and speckle throughputs $\tau_{pla}$, $\tau_{zod}$, and $\tau_{spe}$, and inner and outer working angles $\Theta_{IWA}$ and $\Theta_{OWA}$. The coronagraph can either channel light directly to the imager or through an IFS, which disperses the light with spectral resolution $R$. In both cases, the light is directed toward an EMCCD with effective quantum efficiency $\eta$, dark current count rate of $i_d$ per pixel per second, read noise $i_r$ per pixel per frame, and noise related to clock-induced charge $q_{cic}$ per pixel per frame. Fiducial values for the two *WFIRST*-CGI coronagraphs and the *WFIRST*-CGI EMCCD in photon-counting mode are included in the Appendix in Table 1. For this work, we assume that spectra and images will both be taken in photon-counting mode.

Further detail on the EMCCD and the meaning of the photon-counting mode can be found in the Appendix.

We pause here to emphasize the wavelength-dependent nature of the coronagraph and detector. For a given coronagraph design $\Theta_{IWA}$ and $\Theta_{OWA}$ are fixed in units of wavelength over telescope diameter. The target system's planet–star separation and distance from observer set its working angle. As images are taken with shorter and shorter wavelengths, the coronagraph's $\Theta_{OWA}$ may move interior to the fixed planet image. Alternatively, as images are taken with longer and longer wavelengths, the coronagraph's $\Theta_{IWA}$ may move exterior to the fixed planet image. Similarly, because $C_{raw}$ and throughput depend on working angle in units of wavelength over telescope diameter, the target will experience different coronagraph performance as a function of wavelength (see Figure 1). Quantum efficiencies are also wavelength dependent due to the nature of the silicon and coatings used to make CCDs. Dark current, read noise, and clock-induced charge do not directly depend on wavelength but the PSF-size and thus the number of pixels containing planet signal will scale with wavelength, incorporating more or less detector noise accordingly.

The central quantity under consideration for this model is the S/N:

$$S/N \equiv \frac{Signal}{\sigma_{tot}}. \quad (6)$$

The signal is defined as:

$$Signal = r_{pl}\Delta t, \quad (7)$$

where $r_{pl}$ is the average count rate due to photons arriving from the planet and $\Delta t$ is the total exposure time spent on target (not including calibration time spent observing nearby star and brighter reference star). $r_{pl}$ will depend on the brightness and distance to the star observed, the separation between planet and host star, planet size, albedo, and scattering properties, and the phase angle of the planet at the time of observation. The contrast ratio for a Jupiter–Sun twin system would be $\sim 7.0 \times 10^{-10}$ at quadrature, assuming a Lambertian phase function. When observed from a distance of 10 pc, this gives a rate of $\sim 280$ photons hr$^{-1}$ hitting the telescope at 0.66 microns within a $R = 50$ spectral element (width of 0.0132 microns). However, once the losses due to coronagraph throughput and quantum efficiency are applied, this reduces to a signal rate of only $\sim 1.0\ e^-$ hr$^{-1}$. The noise is defined as:

$$\sigma_{tot} = \sqrt{r_n\Delta t + \sigma_s^2}, \quad (8)$$

where $r_n$ is the total count rate from all random noise sources, and $\sigma_s$ represents an additional systematic noise term accounting for the spatial confusion that arises due to speckle variation over time (Krist et al. 2016; Lupu et al. 2016, and Nemati et al. 2017). The speckle background contributes to both the random noise term $r_n$, and to the systematic noise term. This systematic noise term is defined as:

$$\sigma_s = f_{pp}\, r_{sp}\Delta t, \quad (9)$$

where $f_{pp}$ is a number typically between 1/20 and 1/12 representing both the capabilities of post-processing to remove speckles and the fractional speckle instability inherent to the instrument (WFIRST-CGI engineering team 2018, private communication). Conservatively assuming a value of 1/12 and a $\sim$ Sun-like star at a distance of 10 pc, we get an effective spatial noise term of $\sim 0.2\ e^-$ hr$^{-1}$ at 0.66 microns within a $R = 50$ spectral element after accounting for all losses. Nemati et al. (2017) emphasize that $f_{pp}$ is, at this time, not well-understood in the context of WFIRST-CGI. Proper treatment in future works must consider the dependence on working angle, exposure times, and target and calibration star magnitudes. Despite being poorly understood, it is necessary to include this term because speckles within the dark hole of the coronagraph can mimic planet signals in brightness and spatial extent.

The total noise count rate, $r_n$, is the summation of Poisson noise from photon arrival statistics from both the target and the sky backgrounds, along with detector backgrounds, combined with prefactors 1.32 and 1.20 corresponding to the assumption of background subtraction using RDI against a comparison star that is three magnitudes brighter than the target:

$$r_n = r_{pl} + 1.32(r_{sp} + r_z + r_{ez}) + 1.20(r_d + r_{cic} + r_r). \quad (10)$$

Specifically, $r_{pl}$ is the average count rate due to photons arriving from the planet, $r_r$ is the average count rate originating from read noise, $r_{sp}$ is the average count rate due to photons in the speckle background, $r_z$ is the average count rate originating from zodiacal dust, $r_{ez}$ is the average count rate originating from exozodiacal dust, $r_d$ is the average count rate originating from dark current, and $r_{cic}$ is the average rate of electrons originating from clock-induced charge. To understand the presence of 1.32 and 1.20 in our summation, take Nemati et al.'s (2017) example of a 5th magnitude target star and a 2nd magnitude comparison star. The brightness ratio will be $10^{(5-2)/2.5}$ or 15.8. This allows us to shorten exposure times for calibration stars, thereby reducing detector related noise and ultimately achieving a higher S/N than if we used a comparison star of equal brightness. If we spend 20% of the target time on the comparison star, then the ratio of comparison star variance to target star variance will be $\frac{1}{0.2 \times 15.8}$ or 0.32. In all our calculations, we thus implicitly assume that a comparison star three magnitudes brighter than the target star was observed for 20% of the time on target. This includes the error from RDI in our S/N, while still keeping all of our calculations purely in terms of time on target, $\Delta t$.

S/N calculations depend on the specifics of the planet–star system being observed. Consequently, to provide a demonstration of our model, we have arbitrarily adopted a fiducial target system that falls within the working angles of both coronagraphs for the full wavelength coverage and which has similar physical characteristics to some known RV planets. We fix the 5.0 V-band absolute magnitude star of type G0V, and a one Jupiter radius ($R_J$) planet at an orbital distance of 3.8 au, all observed from a distance of 10 pc. The performance of the EMCCD will degrade as the mission progresses due to radiation damage. For our work, we assume a detector performance consistent with damage incurred from three months at L2 (these values are listed in Table 1). To highlight the relative contributions from the different sources, we fix the product of the planet's geometric albedo and phase function to 0.25 for all wavelengths. Assuming that this fiducial system and detector performance correspond to three months at L2, Figure 3 illustrates the relative count rates from all contributing signal and noise sources for both the IFS and for imaging. Looking at this figure, it is immediately clear that noise arising from the dark current dominates $r_n$ for the IFS with the fiducial values for $i_d$, $i_r$, and $q_{cic}$, and these planet–star system characteristics. The dark current is particularly detrimental to the performance of the IFS in comparison to the imager because the IFS incorporates $\sim 4$ times as many pixels in a single wavelength bin. Meanwhile, $r_n$ for imaging bands is mostly dominated by clock-induced charge because the frame time for imaging used in our model is only 10 s, followed by Poisson noise from the target itself and by zodiacal light at shorter wavelengths but by dark current at longer wavelengths. Again, these calculations are for our fiducial system in particular. In practice, the relative contributions from these sources will depend on the specific characteristics of the target system. For example, a target system with a brighter host star will have a larger contribution from speckle light, exozodiacal light, and planet light relative to the dark current and clock-induced charge counts, or a system with more exozodis of dust will have a larger noise contribution from exozodiacal light. That being said, the trend whereby dark current and clock-induced charge are the dominant noise terms holds for most of the known RV planets that constitute likely *WFIRST*-CGI targets.

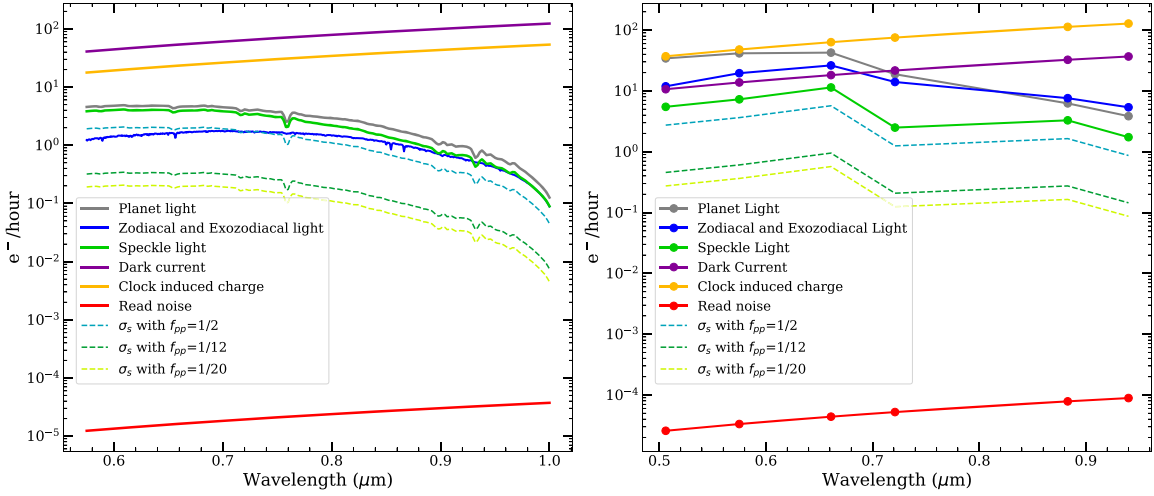

**Figure 3.** Example comparison of the count rates for all sources contributing to signal and noise calculations in the IFS (left) and imaging (right) modes of *WFIRST*-CGI. In both cases, we have assumed our fiducial planet–star system: a 5.0 mag G0V primary and a 1 $R_J$ planet orbiting at a separation of 3.8 au, all observed from a distance of 10 pc. To more clearly show the dependence of noise sources with wavelength, we assume a gray planet and phase function such that $\Phi(\alpha)A_g(\lambda) = 0.25$. In the imaging mode, dots mark the center of each bandpass; we have simply connected them to guide the eye. Spectral dependencies (particularly the dip at $\sim 0.76\ \mu m$) in the planet light, speckle light, and zodiacal light arise from the G0V spectra, taken from the Bruzual–Persson–Gunn–Stryker Stellar Spectrophotometric Atlas, which is distributed with the *HST* Synphot software (http://pysynphot.readthedocs.io/en/latest/appendixa.html). Zodiacal count rates also make use of the solar spectra of Kurucz (2005). The decline in count rates for planet, speckle, and zodiacal sources, beginning at $\sim 0.7\ \mu m$ are consequences of the decreasing quantum efficiency of the detector. In contrast, the dark current and clock-induced charge count rates rise with wavelength as the ratio $\frac{\lambda}{D}$ increases and more pixels are included in the PSF region. In imaging mode, the dark current and clock-induced charge count rates fall nearer the planet light and zodiacal light due to lower resolution and because the imaging signal region incorporates $\sim 4$ times fewer pixels than the IFS signal region. The clock-induced charge has a relatively higher rate for the imaging mode because the frame time will be much shorter in imaging mode to mitigate cosmic-ray hits.

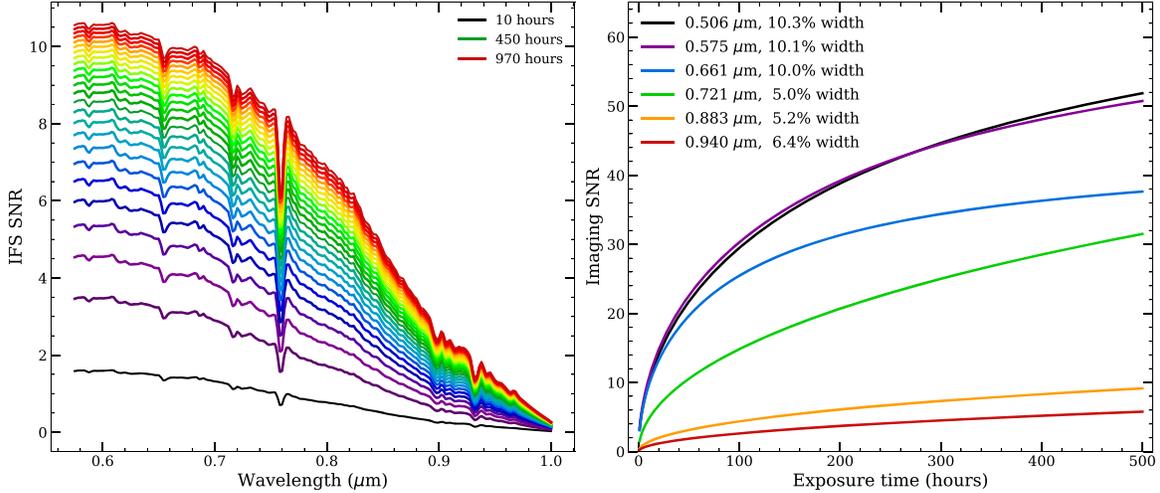

**Figure 4.** The left-hand panel shows S/N achieved with the IFS for on-target exposure times varying from 10 to 1000 hr in 40 hr intervals. At longer wavelengths where larger PSFs include more pixels (thus including more reduced charge and dark current noise), and the quantum efficiency has dropped (reducing the planet signal), the S/N is low even for the longest exposure time of 1000 hr. We have assumed the same fiducial system as Figure 3: a Sun-like star with a Jupiter-sized planet orbiting at 3.8 au observed from a distance of 10 pc. We assume a gray planet and phase function such that $\Phi(\alpha)A_g(\lambda) = 0.25$ to emphasize the wavelength dependence of coronagraph performance rather than the wavelength dependence of the planet atmosphere. Similar to Figure 3, the spectral features seen in the IFS performance originate from the assumed primary star and solar spectrum. The right-hand panel shows the scaling of S/N with exposure time for the six originally planned *WFIRST*-CGI imaging bands using the same fudical system. As is implied by Equation (11), S/N scales approximately with the inverse square root of exposure time (the same is true of the IFS but is less clear in the form plotted). The primary differences between the S/N calculations for images and spectra are the use of the HLC rather than the SPC, and the significantly smaller number of pixels involved in the signal region. Typically, imaging requires an order of magnitude shorter exposure time to achieve any given S/N. The maximum possible values that the S/N vs. exposure time curves will eventually asymptote to depend on the characteristics of the planet–star system being observed—particularly where the planet's working angle falls on the coronagraph because this determines the speckle count rate.

Figure 4 shows the S/N achieved across the IFS coverage and in the six imaging bands for exposure times ranging from 10 to 1000 hr for the IFS and from 0 to 500 hr for imaging for the same fiducial system.

By substituting our definitions of *Signal* and $\sigma_{tot}$ into Equation (6) and re-arranging, we obtain the following expression for $\Delta t$ as a function of S/N:

$$\Delta t = \frac{\mathrm{S/N}^2 r_n}{r_{pl}^2 - \mathrm{S/N}^2 f_{pp}^2 r_{sp}^2}. \quad (11)$$

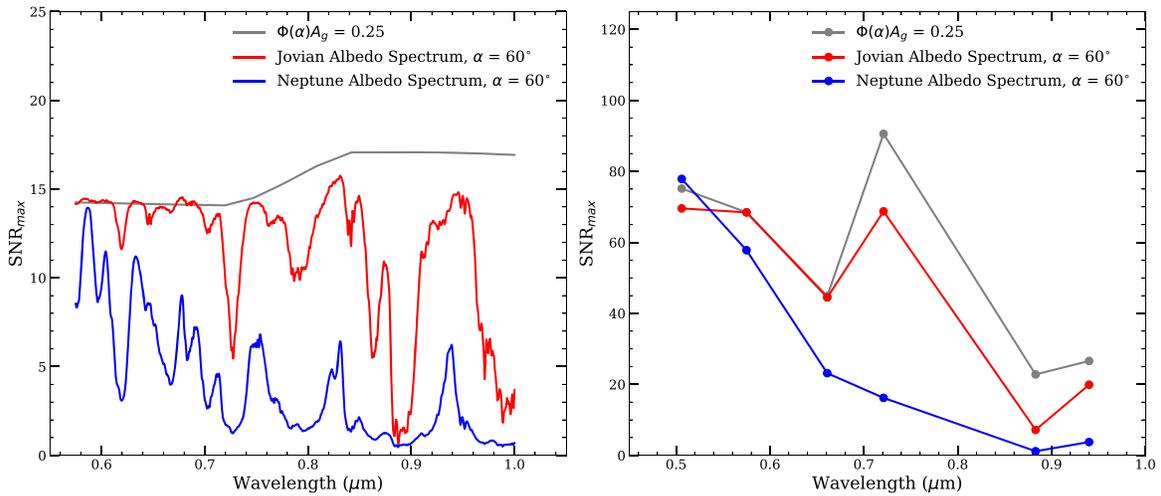

**Figure 5.** Comparison of S/N$_{max}$ as discussed in Equation (12) for three different geometric albedo spectra across the wavelengths in which *WFIRST*-CGI will observe. The left-hand panel shows the IFS capabilities, the right-hand panel shows the imaging capabilities with the six originally planned band centers marked by dots. In both cases, we have assumed our fiducial system: a 5.0 *V*-magnitude G0V type host star with a 1.0 $R_J$ planet orbiting at a separation of 3.8 au, all observed from a distance of 10 pc. In this figure and all calculations in this paper, we assumed a post-processing factor of $f_{pp} = \frac{1}{12}$. The blue and red lines correspond to a planet with the observed geometric albedo spectra of Neptune and Jupiter, respectively (Karkoschka 1994) observed at $\alpha = 65°$, assuming an empirically measured phase function (Mayorga et al. 2016). The gray line corresponds to a gray geometric albedo spectrum such that $\Phi(\alpha)A_g(\lambda) = 0.25$. Achievable S/N levels are about four times higher than the IFS for the imaging bands at wavelengths less than 0.8 $\mu$m, but around the same level as the IFS for the imaging bands at 0.883 and 0.94 $\mu$m. These S/N$_{max}$ values show that it may not be possible to attain high signal-to-noise ratio measurements at the bottom of deep absorption features, regardless of integration time.

Because the denominator includes a difference, if $S/N = \frac{r_{pl}}{f_{pp}r_{sp}}$, then the exposure time is infinite. We thus determine that there is a maximum S/N for a given set of planet and instrument parameters which cannot be exceeded even with infinite integration time. Nemati et al. (2017) refers to this as the critical S/N:

$$S/N_{max} = \frac{r_{pl}}{f_{pp}\,r_{sp}}. \tag{12}$$

For a given $f_{pp}$, if the ratio of the planet light to the scattered speckle light is too low, then S/N$_{max}$ will be less than one, signaling a non-detection of the target regardless of exposure time. In other words, $f_{pp}$ sets a floor which the ratio of planet to speckle light must exceed in order to carry out meaningful observations.

Figure 5 shows S/N$_{max}$ as a function of wavelength for our fiducial planet–star system for three pairings of geometric albedo spectra and phase function. These plots show that, in some cases, this critical S/N may limit our ability to accurately measure the depth of absorption features, restricting the measurement of precise abundances, while perhaps still allowing detection of the *presence* of molecular species (such as methane).

The *WFIRST*-CGI team is working hard to develop new techniques which may render some of these assumptions obsolete and bring about generally more favorable data quantity and quality. The most promising among these are: matched filter spectral extraction which eliminates the need for a comparison star (Kasdin et al. 2003), methods of recovering signal, even with cosmic-ray hits which could allow longer exposures (reducing clock-induced charge), raising core throughput of coronagraph designs, and improving capabilities of deformable mirrors to provide greater speckle stability (essentially lowering $f_{pp}$ in our model).

## 4. Planet Models

### 4.1. Jupiter–Neptune Hybrid Model

In this section, we present a straightforward model for the planet–star flux ratio spectrum of a mature, wide-separation EGP. The optical geometric albedo spectrum will be determined by two parameters representing the atmospheric properties of the EGP, a phase curve will be adopted from empirical observations of Jupiter, and the planetary radius and orbital distance will then determine the observed planet–star flux ratio spectrum as given by Equation (1).

To model the exoplanetary geometric albedo spectrum, we postulate that the atmospheres of Jupiter and Neptune are two extremes of what the atmosphere of an unfamiliar exoplanet might resemble. With this assumption, we can construct a geometric albedo spectrum for a Jupiter-sized planet whose atmospheric properties lie between these extremes. There are a number of contrasting elements between these two giants' atmospheres that motivate this approach, including the order of magnitude difference in metallicity, with methane roughly twenty times more prevalent on Neptune than on Jupiter (Thorngren et al. 2016), the presence of a chromophore of unknown provenance absorbing in the blue portion Jupiter's spectrum, and the significant difference in incident solar flux (Neptune is at ∼30 au, as compared to Jupiter's ∼5 au) that leads to varying cloud-deck depths.

While both planets' reflective properties are dominated by the presence of ammonia clouds, the differences in the albedo spectra of Jupiter and Neptune are largely determined redward of 0.6 $\mu$m by the abundance of gaseous methane in the atmosphere and blueward of 0.6 $\mu$m by the presence or absence of the chromophore. The lower effective temperature of Neptune's atmosphere also plays a role in producing the lower albedo redward of 0.6 $\mu$m because it lowers the ammonia cloud depth. Reflected light must then traverse a larger column depth before emerging from the atmosphere. Because there is no well-established correlation between methane and Jupiter's chromophore, interpolation between Jovian and Neptunian properties is done independently for these two regions. Specifically, for a chromophore-region Jovian character $P_c$ and methane-region Jovian character $P_m$ (i.e., $P_c$ or $P_m = 1$ is Jovian and $P_c$ or $P_m = 0$ is Neptunian in the region $c$ or $m$, respectively), we interpolate the geometric albedo spectra of Jupiter and Neptune as follows:

$$A_g(\lambda) = \begin{cases} A_{g,\text{Jup}} \cdot P_c + A_{g,\text{Nep}} \cdot (1 - P_c), & \lambda < 0.55 \text{ μm} \\ A_{g,\text{Jup}} \cdot P_m + A_{g,\text{Nep}} \cdot (1 - P_m). & \lambda > 0.65 \text{ μm.} \end{cases}$$

$$(13)$$

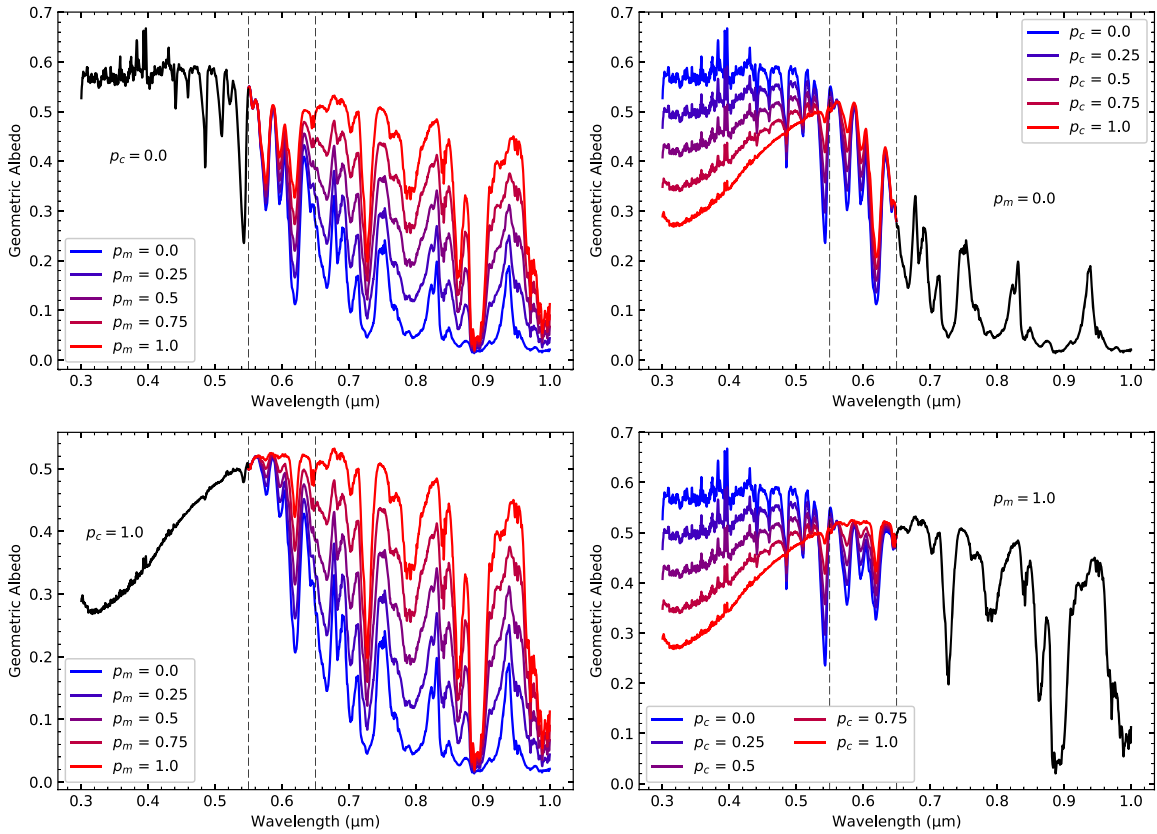

**Figure 6.** Piece-wise interpolations of the geometric albedo spectra of Jupiter and Neptune with a transition region between 0.55 and 0.65 μm. Left: Blueward of 0.55 μm, the spectrum is fully Neptunian (top) or Jovian (bottom), while the methane-dominated region is varied between Jovian ($P_m = 1.0$) and Neptunian ($P_m = 0.0$) conditions. Right: Redward of 0.65 μm, the spectrum is fully Neptunian (top) or Jovian (bottom), while the chromophore-dominated region is varied between Jovian ($P_c = 1.0$) and Neptunian ($P_c = 0.0$) conditions. The left-hand column shows that redward of ∼0.6 μm, the abundance of methane largely determines the geometric albedo of the planet. When the hybrid is closer to having a Neptunian atmosphere ($P_m = 0.0$), its albedo spectrum is lower (particularly note the sharp dips at 0.62, 0.72, 0.79, 0.86, 0.89, and 0.99 μm associated with methane absorption). Neptune's higher methane abundance combines with a lower ammonia cloud depth in its cooler atmosphere to make methane absorption even more prominent in its geometric albedo spectrum. The right-hand column shows that blueward of ∼0.6 μm, the albedo is instead determined by the absorption of a chromophore whose abundance is maximized for hybrid planets with Jovian character ($P_c = 1.0$).

We arbitrarily select the region between 0.55 and 0.65 μm to transition from the blue chromophore-dominated region to the red methane-dominated region using a linearly scaling weighted average of the two formulae provided in Equation (13). Figure 6 shows the resulting albedo spectra for a variety of values of the parameters $P_c$ and $P_m$.

By separating the Jovian character of the red and blue regions of the spectrum, we allow $P_c$ and $P_m$ to function loosely as metrics of the chromophore and methane content of an atmosphere, ignoring the effect of varying cloud-deck depth. In the event that *WFIRST*-CGI only characterizes a few planets, this model could distinguish a Jupiter-like atmosphere from a

Neptune-like atmosphere. If a larger number of targets are characterized, then fits to this model can provide information about trends in methane and chromophore abundances across planets of varying mass and orbital distance, as well as across planets orbiting a variety of stellar types. These trends may then be interpreted to shed some light on the nature and origin of Jupiter's chromophore and the determinants of metal abundances in planetary atmospheres.

Once the geometric albedo spectrum has been established for some hybrid of Jupiter and Neptune, the planet–star flux ratio spectrum for an EGP of known radius and Keplerian elements can be determined by applying the classical phase function (see Equation (1) and Section 2.2). Ideally we would have an empirical phase curve for Jupiter and Neptune with a wavelength dependence resolved on the order with which the *WFIRST*-CGI IFS observations will be done. However, observational data on the phase curves of Jupiter and Neptune are limited. Early reports provided only one phase curve to roughly represent the behaviors of Jupiter, Saturn, Uranus, and Neptune (Pace 1971). More recent measurements from *Cassini* have provided phase curves of Jupiter between $\alpha = 0°$ and $\alpha \approx 140°$ in three optical bandpasses. These data are extrapolated to all phase angles by Mayorga et al. (2016), but they still fail to account for the variation of phase curves over small wavelength intervals. Theoretical models for exoplanetary phase curves (Dyudina et al. 2005, 2016; Cahoy et al.

2010; Hu et al. 2015) limit their scope to exclude Neptune-like planets and they also lack a precise wavelength dependence. Of these limited options, we select, for the hybrid model, the red-bandpass empirical Jovian phase curve from the most recent observations (Mayorga et al. 2016) to represent the true phase curve of our fiducial EGP at all wavelengths. In Section 5, we compare our results derived under this empirical phase curve to those that would be produced under a Lambertian phase curve. Although the Lambertian assumption has been shown to overestimate reflectivities at most phase angles (Greco & Burrows 2015; Mayorga et al. 2016), the prominence of Lambertian phase curves in recent literature motivate this comparison.

In this paper, we also seek to understand the degree to which uncertainties in the wavelength dependence of EGP phase curves will affect characterization of their atmospheres (see Section 5). A wavelength-dependent geometric albedo spectrum. This requires a wavelength-dependent phase curve to pair with our geometric albedo models, which are taken at full phase. Since high-resolution wavelength-dependent empirical phase curves are not yet available, we implement a phase curve that is representative of a uniform atmosphere dominated by Rayleigh scattering. To produce the phase curves corresponding to Rayleigh-scattering atmospheres, we employ the methodology presented by Madhusudhan & Burrows (2012)[5] to generate a phase curve for an EGP with any given albedo spectrum at any given wavelength. By itself, this formalism produces a geometric albedo and phase curve for an atmosphere dominated by Rayleigh scattering under the assumption of a constant single-scattering albedo representative of a uniform atmosphere. Under this formalism, varying the scattering albedo—the ratio of scattering cross section to the summed scattering and absorption cross sections—will necessarily and monotonically vary the resulting geometric albedo and phase curve. We utilize the one-to-one mapping between scattering and geometric albedos to invert a known geometric albedo spectrum into a scattering albedo spectrum, which can then be mapped analytically to a suite of wavelength-dependent phase curves with the same resolution as the input geometric albedo spectrum. Realistically, the assumptions of uniform scattering type and the resulting ability to map a geometric albedo to a unique phase curve might not hold true, but the phase curves resulting from this methodology approximate Jupiter's empirical phase curve fairly well. Figure 7 compares the range of phase curves produced in this manner with a Lambertian phase curve and with Jupiter's phase curve from Mayorga et al. (2016).

The methodology presented in Section 2.2 can be used in tandem with the albedo and phase curve models of this section to generate light curves through *WFIRST* filters for an arbitrary EGP. To do this, we generate an entire spectrum of planet–star flux ratios at each time $t$, combine that with the stellar spectrum to produce a spectrum of exoplanetary flux, and take the ratio of filter-integrated planet flux to filter-integrated star flux:

$$\frac{F_p}{F_*}(t) = \frac{\int_{-\infty}^{\infty} F_p(t, \lambda) \cdot T(\lambda) d\lambda}{\int_{-\infty}^{\infty} F_*(\lambda) \cdot T(\lambda) d\lambda}, \quad (14)$$

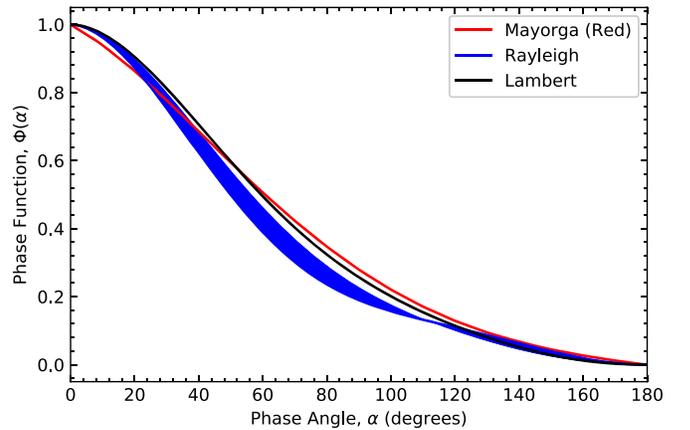

**Figure 7.** Empirical phase curve of Jupiter collected through *Cassini*'s red filter (red curve, Mayorga et al. 2016), compared with the Lambertian phase curve (black) and a suite of wavelength-dependent phase curves for a uniform atmosphere with purely Rayleigh scattering (blue). Note that empirical phase curves through other filters do not vary significantly from the one presented here. The suite of Rayleigh-scattering curves was generated analytically using the formalism of Madhusudhan & Burrows (2012)—for any given wavelength, the resulting phase curve will be contained within the bounded blue region. The bounds correspond to the phase curves associated with the lowest and highest geometric albedos in Jupiter's spectrum, which are roughly consistent with the extreme albedos of all hybrid spectra.

where $t$ (as previously) represents the time elapsed since periastron and the filter transmission $T(\lambda)$ defines the bandpass through which the light curve is generated. Figure 8 compares the resulting light curves for Jovian, Neptune-like, and hybrid planets in various orbits through each of the filters that were originally planned to be incorporated on *WFIRST*.

### 4.2. CoolTLUSTY: Varying Metallicity

In addition to the Jupiter/Neptune hybrid model, we also generated reflection spectra and albedos using the atmosphere and spectral code *CoolTLusty* (Hubeny & Lanz 1995; Burrows et al. 2004; Sudarsky et al. 2005). This code solves self-consistent atmospheres under stellar irradiation by using a detailed suite of thermochemical and opacity data (Burrows & Sharp 1999, augmented to incorporate the ExoMol methane opacities of Yurchenko & Tennyson 2014). However, for albedo calculations for planets with cool atmospheres, once the assumption of chemical equilibrium is made and the atmospheric temperatures are below $\sim$300 K, then variations in the atmospheric temperatures with altitude do not translate into significant compositional or opacity variations. The atmospheres for such giant exoplanets will be dominated by molecular hydrogen and helium (presumably in solar ratios) and gaseous methane. As seen in the atmospheres of Jupiter and Saturn, ammonia is likely to be mostly in condensed phase and constitutes the bulk of the scattering clouds. Some water in vapor and cloud form could also be present, as could other condensates (such as the chromophores seen in the atmospheres of Jupiter and Saturn) or tholins, such as those inferred in Titan's atmosphere. Whatever the condensates (clouds and hazes), they are not at present well-modeled by exoplanet theorists. We do not really know what the species are, nor do we know their spatial (vertical and horizontal) variations or their particle sizes and shapes. Given this uncertainty, for the purposes of this class of models, we merely keep the temperatures below $\sim$200 K, assuming a uniform scattering cloud and a uniform distribution of an absorber. The scattering cloud in our *CoolTLusty* model has a scattering opacity set above a wavelength 0.84 $\mu$m at a constant 0.002 cm$^2$ g$^{-1}$ and below a wavelength of 0.84 $\mu$m it assumes a $\frac{1}{\lambda^{2.5}}$ behavior. The uniform haze absorber is taken to be Titan tholins (Khare et al. 1984) with an assumed atomic weight of 100, a model particle size of 0.05 $\mu$m, and a number fraction of $3.3 \times 10^{-10}$. Even with such a low abundance, the tholin haze can markedly affect the albedo at short wavelengths and serves as our chromophore. These specific numbers and constituents were chosen to fit Jupiter's albedo spectrum (Karkoschka 1994, 1998) for an atmosphere with a metallicity of 0.5 dex, $\sim$3.16 × solar elemental abundances, insolated with a blackbody solar spectrum at 5777 K. As Figure 9 demonstrates, the fit to Jupiter's albedo spectrum is rather good, though not perfect. With this background model, we then varied only the metallicity to include solar, 10× solar, and 30× solar. In this way, we have generated a simple model suite that crudely captures the possible metallicity (read methane) dependence of these exoplanets' albedos. The other computed models and interpolations between them to approximate Jupiter with 2.0, 5.0 and 20.0 times solar metallicity are also provided on Figure 9. We again pair these models for geometric albedo spectra with the empirical Jovian phase curve of Mayorga et al. (2016) and a Lambertian phase curve to compute planet–star flux ratios.

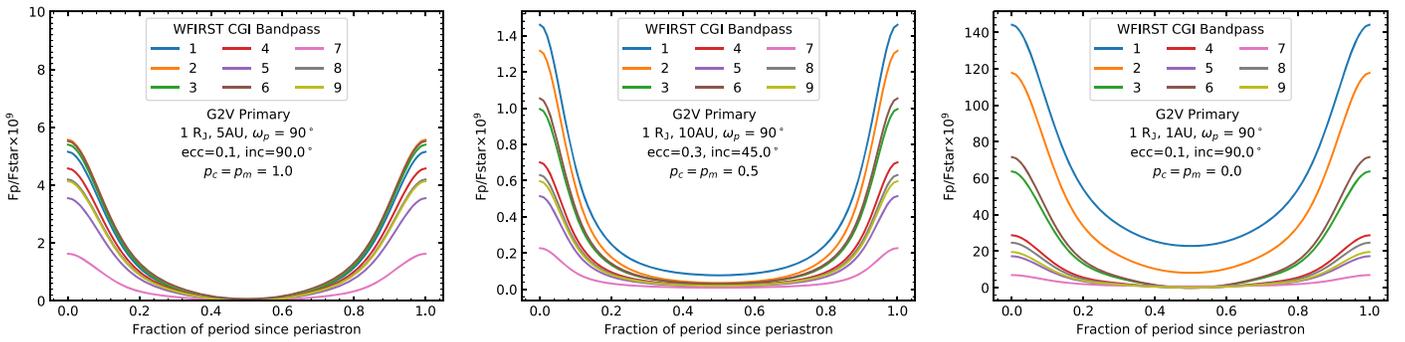

**Figure 8.** Light curves for an assortment of hypothetical EGP's as they would appear through the originally proposed *WFIRST* filters. The left-hand panel shows the light curves generated from a Jovian planet in a low-eccentricity, edge-on orbit 5 au from its parent star. The right-hand panel simulates a Neptunian (but still Jupiter-sized) EGP in a similar orbit just 1 au from the parent star. The central panel displays the light curves for a 50% hybrid-atmosphere planet in a more eccentric orbit, now inclined at 45°, 10 au from the parent star. In all calculations, the argument of periastron $\omega_p$ is assumed to be 90°, and the primary star is assumed to be Sun-like. Note that the light curve through CGI Bandpass 7 (at roughly 0.89 $\mu$m) depicts significantly lower planet–star flux ratios for all of the modeled systems at all times, which corresponds to the strongest methane absorption within the band. In all three cases, variation in the color of the planet with phase angle is apparent.

The *CoolTLusty* approach captures differences across the *WFIRST*-CGI spectral coverage in one physically motivated parameter, which makes it a useful tool for interpreting *WFIRST*-CGI data and other similar observations. Nevertheless, this approach has some drawbacks. In particular, the assumption of known cloud and haze scattering properties could introduce a systematic bias in metallicities fit to EGPs if they deviate significantly from what is expected.

## 5. Results

*WFIRST*-CGI has the potential to observe planet–star systems with a wide array of characteristics. As the signal and instrument models presented in Sections 4 and 3 illustrate, many factors will determine the relative contributions of signal and noise over exposure time, including the distance from the observer, the spectral type and the magnitude of the host star, the planet–star separation, the planet's radius, the planet's wavelength-dependent geometric albedo, the phase angle at the time of observation, the wavelength-dependent performance of the coronagraph, and the observing strategy adopted by the *WFIRST*-CGI mission.

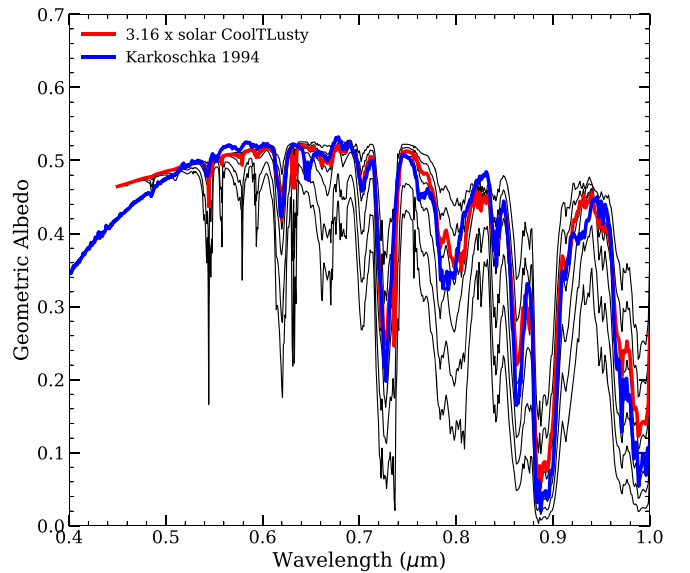

**Figure 9.** Example geometric albedo spectra generated using *CoolTLusty* to first match the measured Jovian spectrum (Karkoschka 1994, bold blue line) at a fixed metallicity of 0.5 dex, $\sim$3.16× solar elemental abundances, (bold red line), and then fixing all cloud, haze, and particle parameters and simply perturbing metallicity (thin black lines show abundances of 1.0, 2.0, 5.0, 10.0, 20.0, and 30.0 times solar metallicity). Although the match between Jupiter's observed spectra and the best fit with *CoolTLusty* is not perfect, it is certainly adequate for the purposes of this study.

To present a realistic and coherent understanding of *WFIRST*-CGI's science potential in the context of our two classes of geometric albedo model, we adopt a fiducial set of planet–star characteristics that are similar to those used in Section 3: a $\sim$Sun-like star with spectral type G0V and an absolute *V*-band magnitude of 5.0, a 1.0 $R_J$ planet, a planet–star separation of 3.8 au, a circular orbit seen edge-on, and observed at $\alpha = 65°$ from a distance of 10 pc. These characteristics are chosen because they keep the target within the functional working angle of both coronagraphs across the full range of spectral coverage, while maintaining consistency with the physical characteristics of detected RV planets deemed suitable targets for *WFIRST*-CGI (see Section 5.5). Note that a planet in this fiducial system only spends $\sim$30% of its orbit at phase angles detectable in reasonable exposure times and within that fraction only 10% of its orbit is near our fiducial phase angle ("near" meaning between $\alpha = 55°$ and $\alpha = 75°$). It is very optimistic to think that the most desirable target will be at a phase angle of precisely 65°. However, given the number of potential known RV planet targets and the likely discovery of additional planets prior to *WFIRST*-CGI's flight, it is not unreasonable to expect that one or two targets will be found at phases that are amenable to observation within the mission's window.

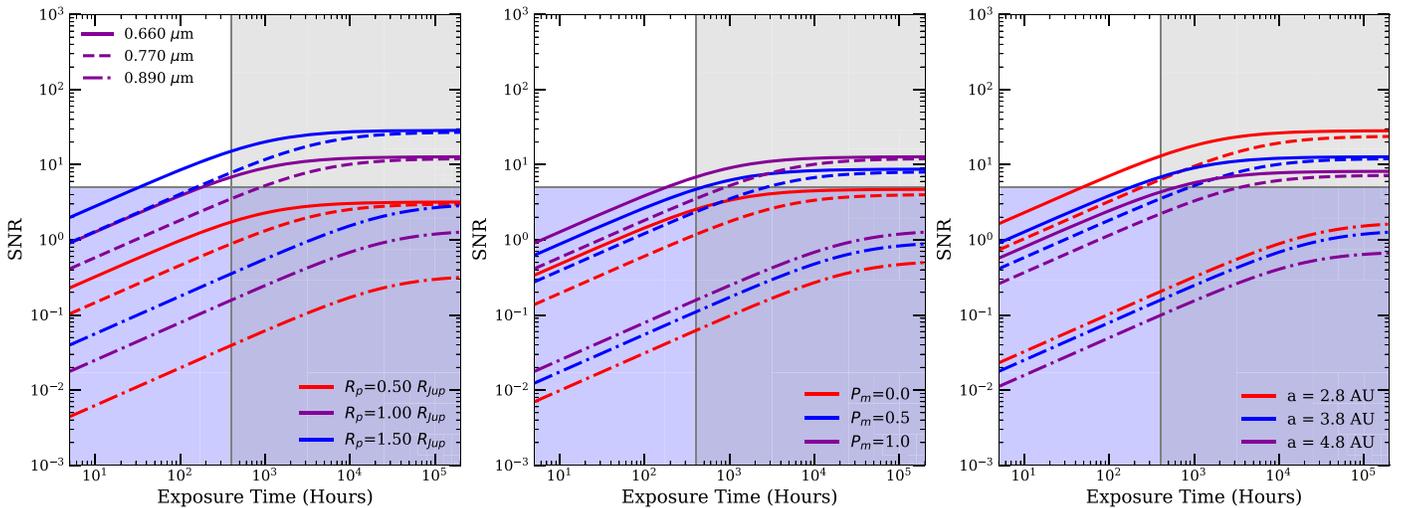

**Figure 10.** S/N as function of exposure time for the $R = 50$ spectral element of the IFS at the three band centers. In all three panels, solid lines correspond to 0.66 $\mu$m, dashed lines correspond to 0.76 $\mu$m, and dashed–dotted lines correspond to 0.89 $\mu$m. Colors indicate a variation in a single parameter from the fiducial star–planet system: a Jupiter-size planet, with a Jovian geometric albedo spectrum, with separation of 3.8 au, orbiting a ~Sun-like star, observed from 10 pc away at $\alpha = 65°$ assuming the observed phase curve of Jupiter (Mayorga et al. 2016). The left-hand panel varies planet radius. The center panel varies the value of the methane-like absorption parameter $P_m$ for the Jupiter–Neptune hybrid model. The right-hand panel varies the semimajor axis of the planet's orbit, which is also the separation because we have assumed circular orbits. Exposure times above 400 hr are shaded gray because they exceed the desirable exposure time for *WFIRST*-CGI. S/N values below 5 are shaded purple because this is a threshold above which previous work has concluded that the spectra are scientifically useful. If we ignore the $\sigma_s$ correction term in Equation (6), then S/N scales roughly with the square root of the exposure time. This shape is apparent in all panels in the ~1/2 slope on the log–log scale. At late times, the S/N asymptotes to the maximum achievable value, which can be seen as the curve flattens off. The S/N's achieved for 0.89 $\mu$m hardly exceed 1 because it falls near the bottom of a strong methane absorption feature and in the region where quantum efficiency has plunged. Increasing planet radius increases S/N across all regions of the spectrum in a similar manner. Lower values of $P_m$ indicate a planet closer to Neptune-level methane absorption and its correspondingly very low albedos, while higher values of $P_m$ indicate a planet closer to Jupiter's methane absorption with slightly higher albedos. Changing the semimajor axis changes both the amount of stellar insolation reaching the planet and the working angle of the planet on the detector.

First, we use this fiducial system to illustrate exposure time and S/N relations in the three originally planned IFS bands and the six originally planned imaging bands. Then, we explore the degree to which four observational scenarios utilizing various combinations of imaging and IFS coverage can constrain our two simple planet albedo models and planet radius. Finally, we set aside this fiducial system and compute achievable S/Ns for known RV planets, selecting two promising target systems (47 UMa c and Upsilon And d) and simulating observations for the same four observational scenarios.

### 5.1. Signal-to-noise Ratios and Exposure Times

Relating S/N to exposure time is an essential part of planning observational strategies with *WFIRST*-CGI. We explored how this relation varies with radius of the target planet, planet–star separation, degree of methane absorption, and phase angle by computing values of S/N for the central spectral element of each IFS wavelength bin (see Figures 10 and 12), and for the six originally planned imaging bandpasses (see Figures 11 and 13), all as a function of on-target exposure time. These figures provide a clear picture of the general scaling of S/N with time for the model presented in Section 3. For low exposure times, S/N increases with the square root of exposure time, exhibiting a slope of 0.5 on our log–log plots. At longer exposure times, S/N asymptotically approaches the maximum value given by Equation (12). Integration times above 400 hr are not desirable for a single IFS bandpass and integration times above 100 hr are not desirable for a single imaging bandpass; therefore, these regions in the figures are shaded gray. Previous works identified an S/N of five as sufficient to gain useful science information with spectra (Lupu et al. 2016), so the region below this threshold is shaded purple. In the case of our fiducial system and these variations of it, the S/N attained is limited by the 400 hr maximum desirable exposure time for the IFS at shorter wavelengths and the 100 hr maximum desirable exposure time for the imagers, rather than the maximum S/N (see how the curves flatten in the gray shaded regions of Figures 10 and 12). Section 5.5 shows similar calculations for a selection of planet–star systems detected with RV.

*WFIRST*-CGI is able to attain an S/N of 4–5 for our fiducial system in under 400 hr with the IFS bandpasses centered at 0.66 and 0.76 $\mu$m (see curves corresponding to $R_p = 1.00R_{Jup}$, $P_m = 1.0$, $a = 3.8$ au, and $\alpha = 65$ degrees, in Figures 10 and 12). In addition, curves for the larger planet radius, smaller planet–star separation, and smaller phase angles demonstrate that some realistic target systems will provide similar or slightly superior data quality relative to our fiducial system in the same exposure time. The central element of the IFS bandpass centered at 0.890 $\mu$m does not ever attain an S/N over 5, assuming that there is methane absorption on the level seen in Jupiter and Neptune's atmospheres. For likely target systems, exposure times will be on the order of several hundred to a thousand hours for the IFS to attain spectra with S/N ~ 5, with exact times depending on the planet–star system characteristics. Portions of the IFS coverage that fall at the bottom of the strong absorption features will be unlikely to attain an S/N over 5, even for infinite exposure time.

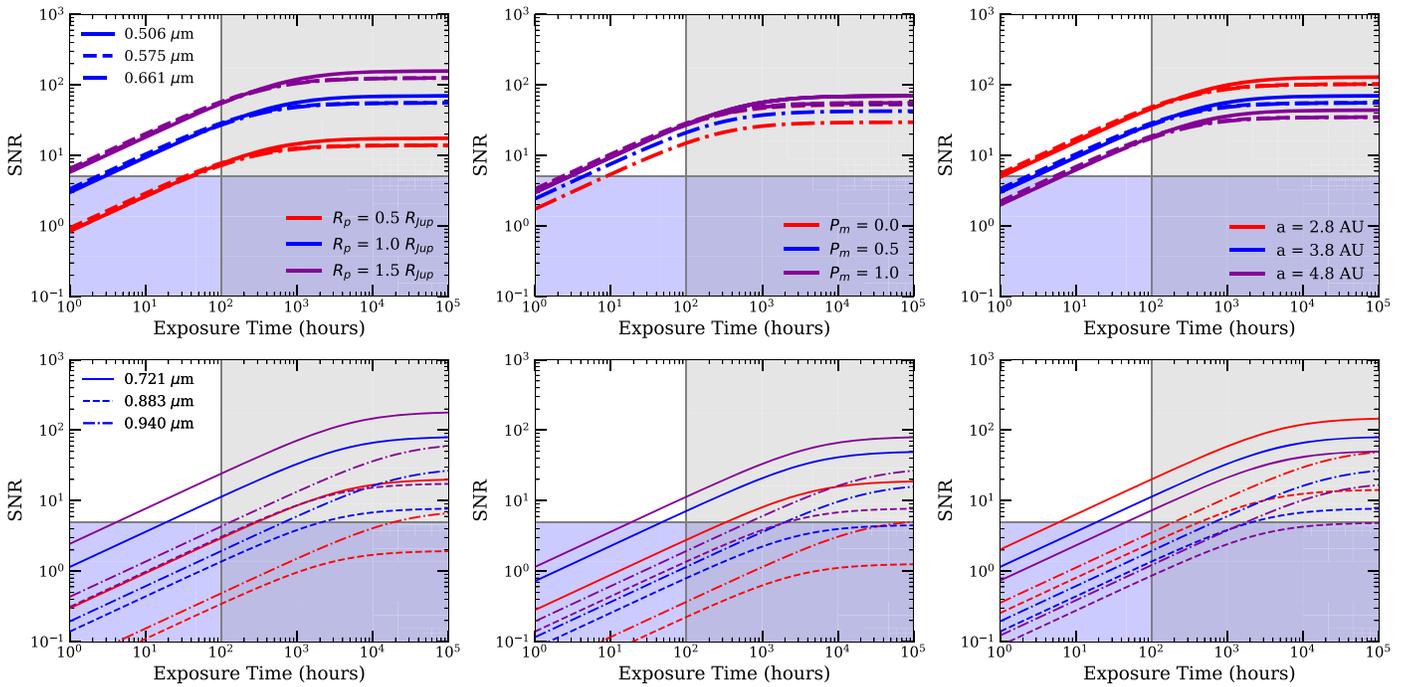

**Figure 11.** S/N as a function of exposure time for the six originally planned imaging bandpasses. The top row shows the 0.506, 0.575, and 0.661 $\mu$m bands as solid, dashed and dashed–dotted lines, respectively. The bottom row shows the 0.721, 0.883, and 0.940 $\mu$m bands, as thinner solid, dashed, and dashed–dotted lines respectively. Colors indicate a variation in radius (left column both top and bottom panels), long-wavelength absorption (center column both top and bottom panels), and semimajor axis (right column both top and bottom panels), with all other planet–star system characteristics held to our fiducial values as defined at the start of Section 5. Gray shading indicates exposure times above 100 hr. Purple shading indicates S/N below five. Similar to the IFS, the imaging S/N scales with the square root of the exposure time at short times, eventually asymptoting to a maximum value at longer times. Aside from the 0.883 $\mu$m band and the 0.94 $\mu$m band which only attains reasonable S/N within 100 hr for the 1.5 $R_J$ radius planet at a separation of 3.8 pc, or for the 1 $R_J$ radius planet at a separation of 2.8 au, the imaging bands all attain desirable S/N levels across a variety of realistic planet radii, methane absorption levels, and planet–star separations.

Imaging in the 0.506, 0.575, 0.661 and 0.721 $\mu$m bands will easily attain S/N's of ∼30 within the maximum imaging time of 100 hr for our fiducial system, while the longer wavelength 0.883 and 0.940 $\mu$m bands will not exceed an S/N of 5 in 100 hr. The near order-of-magnitude variation in S/N for a fixed exposure time between the imaging bands arises because the wavelength dependence of the coronagraph performance, lower typical gas giant albedos, plunging quantum efficiencies, and growing PSF sizes are larger, which conspire to lower the S/N relative to the shorter wavelengths.

These results make it clear that the longest wavelength band for the IFS will only yield useful S/N data under the most favorable planet–star system circumstances (which are not currently met by any of the known RV planets), or in the absence of methane absorption. Even the longest wavelength imaging bands perform significantly worse than the shorter wavelength bands. If a starshade is eventually paired with *WFIRST*, then it will provide significantly higher throughput than the coronagraph and perhaps allow these longer wavelengths to attain reasonable S/N. For now, these wavelengths can be better probed by imaging than by the IFS.

Figures 12 and 13 show, unsurprisingly, that observations should be done at the phase angle that corresponds to maximum illumination while still falling at a working angle exterior to the inner working angle (typically in the range 60°–80°). We assume that a Lambertian phase function results in a higher S/N than the observed phase function of Jupiter for any given exposure time, with the largest difference in the range $\alpha = 50°$ to $\alpha = 90°$, which are the most likely phases that *WFIRST*-CGI will target. This is consistent with the long noted discrepancy between a Lambertian phase function and reality (as shown in Figure 7), and hints at the importance of factoring uncertainties in planet phase and phase function into future analyses. The differences between the two phase functions are less pronounced at $\alpha > 110°$ and $\alpha < 20°$.

We also computed the exposure time necessary to reach an S/N of 5 across the full IFS wavelength coverage, as a function of wavelength and phase, wavelength and exozodi level, wavelength and planet radius, and wavelength and planet–star separation (Figure 14). Integration times were again computed assuming our fiducial planet–star system, aside from the parameter that was being varied. These calculations of exposure time in terms of two parameters make it apparent that only certain portions of the exoplanetary spectrum might attain adequate S/N or even be observable at all depending upon the planet–star system characteristics. Any region in Figure 14 that is darker green, blue, purple, or black is unable to reach an S/N of 5 for our fiducial system with *WFIRST*-CGI's IFS, either due to S/N$_{max}$ limitations or due to exposure times that exceed the mission's capabilities and resources. These regions occur in deep absorption features in both the stellar spectrum and the planet albedo spectrum, below a minimum radius of ∼0.8 $R_J$, and above a maximum phase angle of $\alpha \sim 90°$. White portions of the figures represent areas where the planet will fall outside the inner or outer working angles of the coronagraph (recall $\Theta_{IWA} \sim 3\frac{\lambda}{D}$ and $\Theta_{OWA} \sim 9\frac{\lambda}{D}$ for both HLC and SPC in its "bow-tie" field of view configuration).

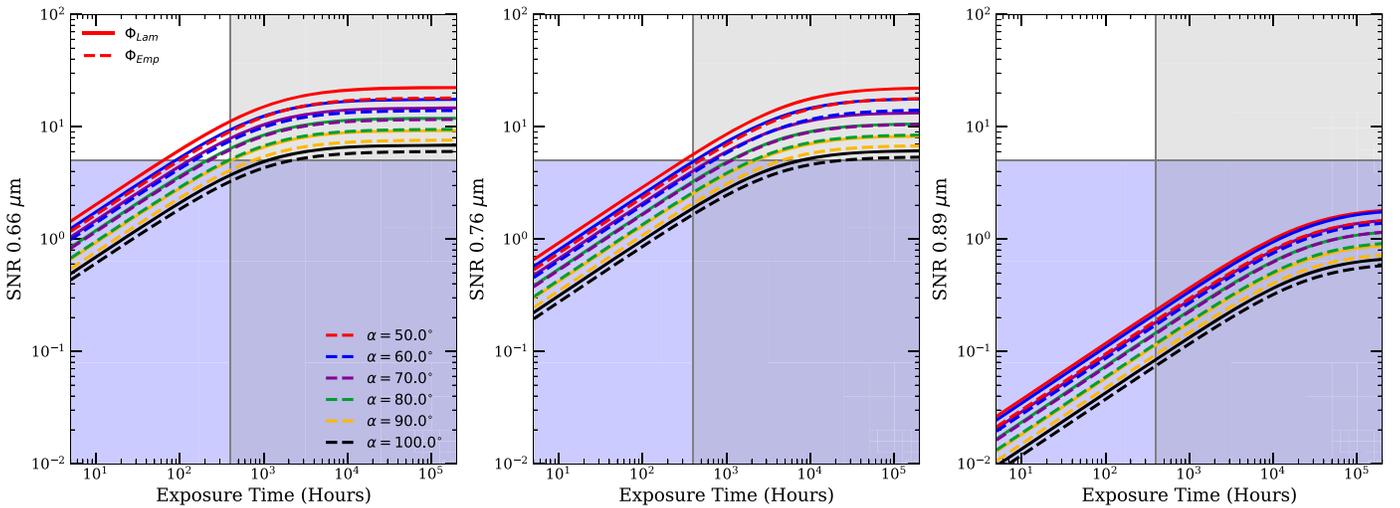

**Figure 12.** S/N as a function of exposure time for the $R = 50$ spectral element of the IFS at the three band centers (0.66 $\mu$m on the left, 0.76 $\mu$m in the center, and 0.89 $\mu$m on the right). Colors indicate the phase angle for which the S/N is calculated, with all other planet–star system characteristics held to our fiducial values defined at the start of Section 5. In all of th panels, solid lines correspond to the Lambertian phase function and dashed lines correspond to the empirical Jovian phase function of Mayorga et al. (2016). Gray shading indicates exposure times above 400 hr, which will not be feasible with *WFIRST*-CGI. Purple shading indicates S/N < 5. As expected, the Lambertian phase curves estimate a higher S/N for any given exposure time than the observed Jovian phase curve. See Figure 7 for a comparison of the Lambertian and empirical phase curves. For our fiducial system, phase angles between 50° and 80° produce S/N > 5 within 400 hr in the 0.66 $\mu$m and 0.76 $\mu$m bands, while larger phase angles do not.

It is not surprising that planet radius, planet–star separation, and phase angle all strongly influence integration times to reach an S/N of five. If we ignore the $\sigma_s$ correction term (combining Equations (6) and (8)) in our model, then the exposure timescales approximately linearly with the planet's signal. This happens because the S/N scales approximately with the square root of the planet count rate, and the exposure time in turn scales as the square of the S/N. The planet's photon count rates scale with radius squared (see Equation (1)), so exposure times will scale with radius squared. Similarly, the planet's count rates scale inversely with the square of planet–star separation, but we must also fold in the effects of varying the working angle of the coronagraph for a fixed observing distance. Varying the phase angle affects the count rates from the planet according to the phase function. Once again, by changing the working angle where the planet falls on the coronagraph, it will move through different levels of coronagraph performance. In contrast, exozodi levels do not have a strong effect. $N_{ez}$ multiplies directly into exozodi count rates but exozodi is not a dominant noise term for the IFS, particularly at a separation of 3.8 au and for a 1.0 $R_J$ planet. Therefore, in the upper right-hand pane we can see that it causes little variation in the requisite exposure times for any given wavelength. This is consistent with the findings of Robinson et al. (2016).

### 5.1.1. Observing Neptune-sized Planets with WFIRST-CGI

What is the smallest target planet that can be seen with *WFIRST*-CGI? As Section 5.1 explains, the ratio of signal and noise from a given planet–star system depends on many factors. Planets that are observable near full illumination and which are closer to their host stars (all while maintaining the high albedos associated with clouds) will be brightest for any given radius. The limiting radius is thus highly dependent on the distance to the target system for three reasons: (1) nearer stars will allow observations of planets with smaller separations, (2) observations can be done closer to full phase, and (3) more photons reach the observer, scaling with the inverse square of distance. We computed integration times to reach an S/N of five for a planet with a radius of 0.4 $R_J$ and a Jovian geometric albedo spectrum around a G0V star with absolute $V$-band magnitude of 5, as a function of wavelength and phase angle (this calculation is similar to the one shown in Figure 14), at a grid of observer distances and planet–star separations. These calculations indicate that *WFIRST*-CGI may obtain spectra in the 0.66 $\mu$m band for a Neptune-sized planet out to 8 pc from Earth if it is at a separation of 1.5–2 au and is observed at phase angles between 40° and 70°. If the planet–star system is instead 9 pc from Earth and the orbital distance is still 1.5–2 au, then it will not attain useful S/Ns with the IFS, although it could be detected by shorter wavelength imaging. Between 8 and 9 pc, IFS observations may be successful— depending on the geometric albedo of the planet. Moving the planet to a wider separation quickly lowers the S/N and lengthens exposure times. Moving it to a smaller separation provides a more favorable signal but requires that the target star be within ~5 pc to fall in the working angle of the coronagraph. As currently planned, observing planets with radii significantly smaller than Neptune could be plausible with the imager, and even then only for *very* nearby stars. Lowering the resolution of the spectrograph could allow observations of smaller planets with the IFS. Overall, we find that *WFIRST*-CGI will have the most success observing roughly Jupiter-sized planets but it will still be able to probe Neptune-sized planets under favorable circumstances.

### 5.2. Simulating Observations and Fitting Procedure

Given our planet models and Nemati et al.'s (2017) model for the *WFIRST*-CGI S/N, it is straightforward to generate a noisy spectrum or photometric measurement given an on-target exposure time. To generate a single realization of a noisy synthetic observation, for each photometric and spectral bin one ought to draw a random value from a normal distribution with the mean being the true value of the spectrum/photometry and the variance being $\sigma_{tot}$, as defined in Equation (8). We

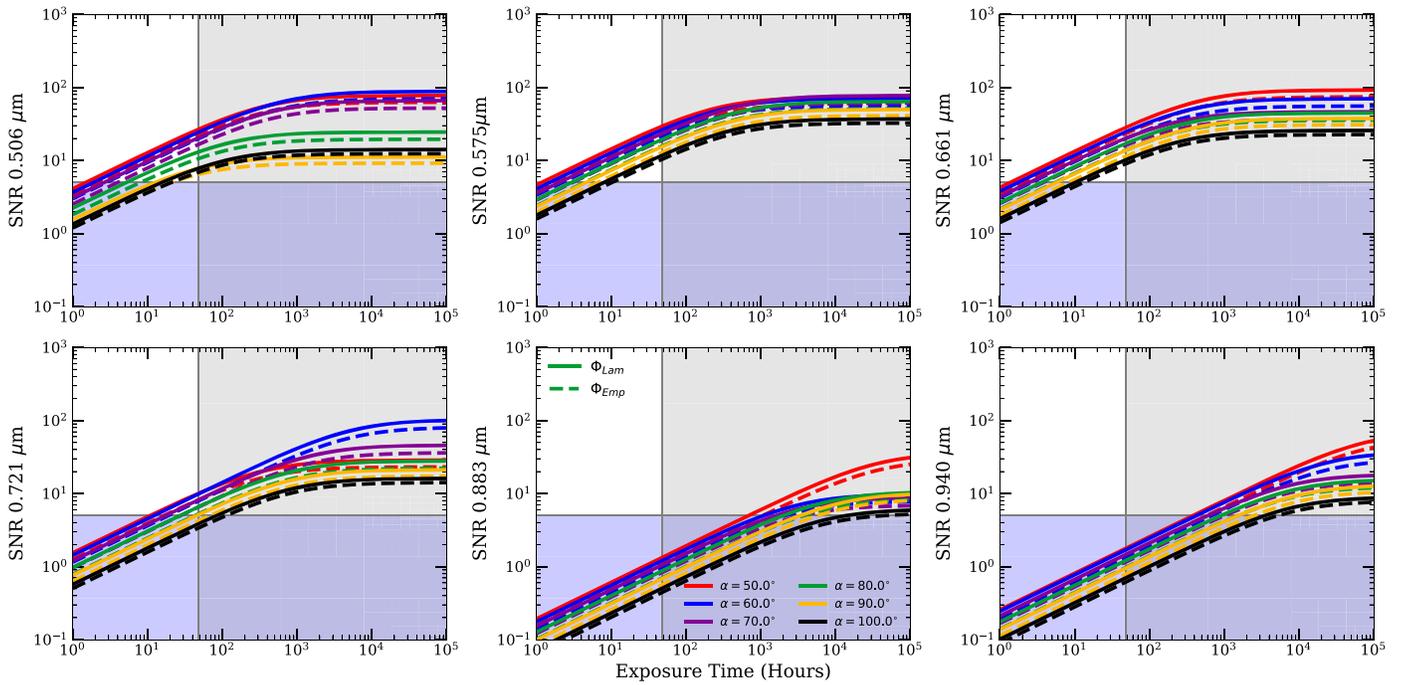

**Figure 13.** S/N as a function of exposure time for the six *WFIRST*-CGI imaging bands (each bandpass shown in a separate panel and labeled with its central wavelength) computed at phase angles from 50° to 100°, as indicated by the line colors. The dashed lines are the empirical Jovian phase function from Mayorga et al. (2016), which are compared to a classical Lambertian phase function that is indicated by a solid line. The gray shaded region again indicates exposure times above 100 hr and the purple shading indicates S/N below five. The differences are much more pronounced at phases where bands fall near the inner or outer working angle of the coronagraph (see Figure 1), leading to sudden changes in coronagraph performance with working angle. As long as the phase angle does not draw the planet outside the peak performing working angles of the coronagraph, imaging S/N is adequate across this range of phase angles for all the bandpasses besides 0.883 and 0.94 μm. As expected, the Lambertian phase function attains a higher S/N than the empirical phase function for a given exposure time.

choose, instead, to use the true clean signal values and assign them the uncertainties $\sigma_{tot}$, which is the result that one would expect if the measurement were made an infinite number of times. We consider four possible observing cases:

Observing Case One: Full IFS coverage across the three ~18% width bands at 0.66, 0.76, and 0.89 μm, the two shortest wavelength imaging bands centered at 0.506 and 0.575 μm, and the imaging band centered at 0.883 μm. This incorporates almost all of the capabilities that the original *WFIRST*-CGI mission would have had. For the sake of simplicity, we take the total exposure time and divide it such that we have the same exposure time for each IFS bandpass, an exposure time that is ~1% the length of the IFS exposure time for the shorter imaging wavelengths, and an exposure time of ~50% the length of the IFS time for the 0.883 μm imaging bandpass. We then compute the signal and noise contributions given the appropriate exposure time for each photometric measurement and in each wavelength bin spanning the three IFS filters with a spectral resolution $R = 50$. Note that this results in slightly higher spectral resolution where the filters overlap.

Observing Case Two: Imaging only, done in all six originally planned bands: 0.506, 0.575, 0.661, 0.721, 0.883, and 0.940 μm. Again, this case would have only been possible with the originally planned *WFIRST*-CGI capabilities. In case two, we divide the total observing time such that the three narrower (5% width) bands at 0.721, 0.883, and 0.940 μm have the same exposure time and the three wider imaging bands at 0.506, 0.575, and 0.661 μm have an exposure time 50% the length of the narrower imaging band exposure time.

Observing Case Three: One ~18% width IFS band centered at 0.76 μm and one imaging band at 0.575 μm. This case only uses the coverage that will be fully commissioned for the technology demonstration *WFIRST*-CGI mission. In this case, we divide the total exposure time such that the imaging band has an exposure time that is 1% the duration of the IFS exposure time.

Observing Case Four: One ~18% width IFS band centered at 0.76 μm, an imaging band at 0.575 μm, and an imaging band at 0.883 μm. This case is meant to *roughly* approximate using all three of the fully commissioned bandpasses. The largest differences are that the fully commissioned 0.825 μm bandpass is locater further from a deep absorption feature and will have double the width of the 0.883 μm bandpass that we use. For simplicity, we also neglect to use the correct SPC configuration, which would alter the inner working angle for the long-wavelength imaging. In this case we once again divide the total exposure time such that the 0.575 μm imaging band has an exposure time that is 1% the duration of the IFS exposure time and the narrower 0.883 μm imaging band has an exposure time that is 50% the duration of the IFS exposure time.

Examples of the four observing cases for our fiducial system are shown in Figure 15. For each observing case, the figure shows the underlying high-resolution planet–star flux ratio spectrum, the regions of the spectrum which each observing case would sample (with corresponding uncertainties calculated using our noise model), and a random selection of models from the posterior distribution mapped by the MCMC chains. Our simplistic division of the total exposure time results in slightly variable S/N for the different photometric measurements. In practice, exposure times would likely be varied to make the most efficient possible observation that still attains an adequate S/N. Again, we emphasize that cases three and four use only bandpasses which approximate bandpasses that will be fully commissioned as part of the mission's requirements, while cases one and two also use some of the bandpasses that were originally planned prior to de-scoping the mission to a technology demonstrator. Under the guideline that IFS exposure times above 400 hr and imaging exposure times above 100 hr are undesirable, we can form a rough idea of the maximum total observing time desirable for each observing case: case one ∼1500 hr, case two ∼600 hr, case three ∼500 hr, and case four ∼600 hr.

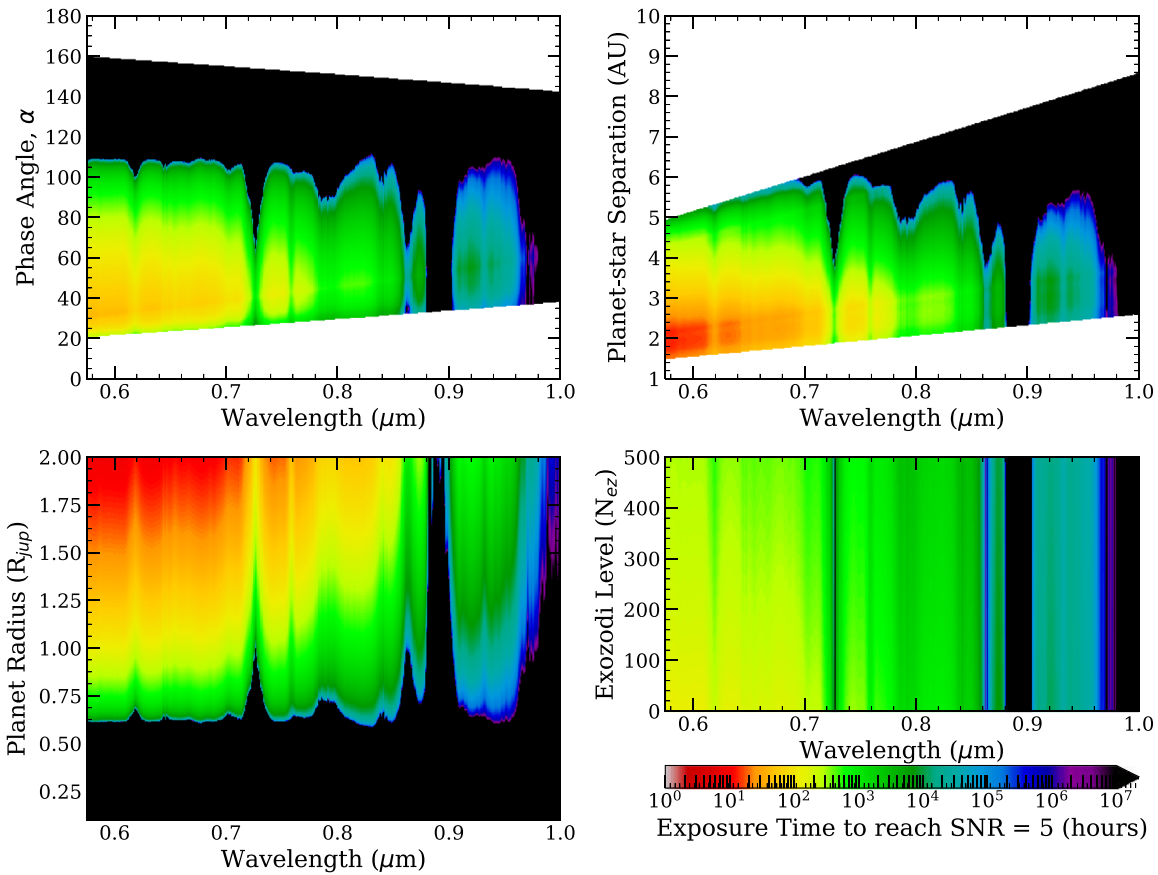

**Figure 14.** Integration time required to reach an S/N of 5 in each spectral element of the IFS as a function of phase angle, exozodi level, planet radius, and planet–star separation vs. wavelength. Integration times were computed assuming our fiducial planet–star system (defined at the start of Section 5) aside from the parameter that was being varied. We assume a Jovian geometric albedo spectra and Mayorga et al.'s (2016) empirical Jovian phase curve. Integration times above 400 hr are not feasible, so any region that is dark green, blue, purple or black can be interpreted as unable to reach an S/N of 5 with *WFIRST*-CGI. Note that some subsections of the black regions have S/N$_{\rm crit}$ < 5, indicating an infinite integration time on this plot. The white space in the upper left and upper right-hand panels represent areas where the planet will fall outside the inner or outer working angle of the coronagraph at that wavelength. These results indicate that for systems like our fiducial system (that is, systems at a distance of 10 pc), observations should be done when planets are at phase angles between 20° and 80°, exozodi does not have a large effect on exposure times to reach S/N of 5, planets smaller than 0.6 $R_J$ will not be observable, and separations of 2–4 au provide the best balance of wavelength coverage and planet signal. Note that planet radii approaching 2.0 $R_J$ are only seen when atmospheres are inflated due to close proximity to the host star. These close-in planets will not fall in the working angles of the coronagraph unless they are around an extremely nearby star. Additionally, they will not have geometric albedo spectra similar to Jupiter.

To investigate the synthetic observations' ability to constrain the planet geometric albedo model parameters and the radius of the planet, we carry out a simultaneous fit of all parameters using *emcee* (Foreman-Mackey et al. 2013), which is a parallelized Python implementation of the affine-invariant MCMC methods presented in Goodman & Weare (2010). We take the median of the resulting samples of the posterior distribution to be the recovered parameter value and compute the 68% confidence interval, or 1σ error on that value.

Adopting our fiducial planet–star system, we carried out this procedure for total integration times ranging between 100 and 10,000 hr for a grid of $P_c$ and $P_m$ values using the hybrid Jupiter–Neptune geometric albedo models, and for a series of metallicities using the *CoolTLusty* model. We also consider a range of phase angles between 50 and 130 degrees and the same range of exposure times, using both a Lambertian phase function and Mayorga et al.'s (2016) empirically measured phase function. A representative selection of these results is shown in Sections 5.3 and 5.4. In addition to these comprehensive explorations of phase and exposure time dependence using our fiducial system, we also consider the systems 47 UMa c and Upsilon And d at two planet model parameterizations, two phases, and two exposure times in all four observing cases. Results are shown in Section 5.5.

As noted previously, these 100 hr to 10,000 hr integration times are the total on-target science exposure time for each observing case. They do not account for the calibrations that would be included in the total on-sky integration time for a whole observing case. It will take time to return the coronagraph to full performance after switching between different imaging and IFS filters, and after switching between the two coronagraph designs intended for use with spectroscopy and the IFS. Consequently, we chose to extend our study to times an order of magnitude above the desirable limit of 1500 hr to more clearly show trends.

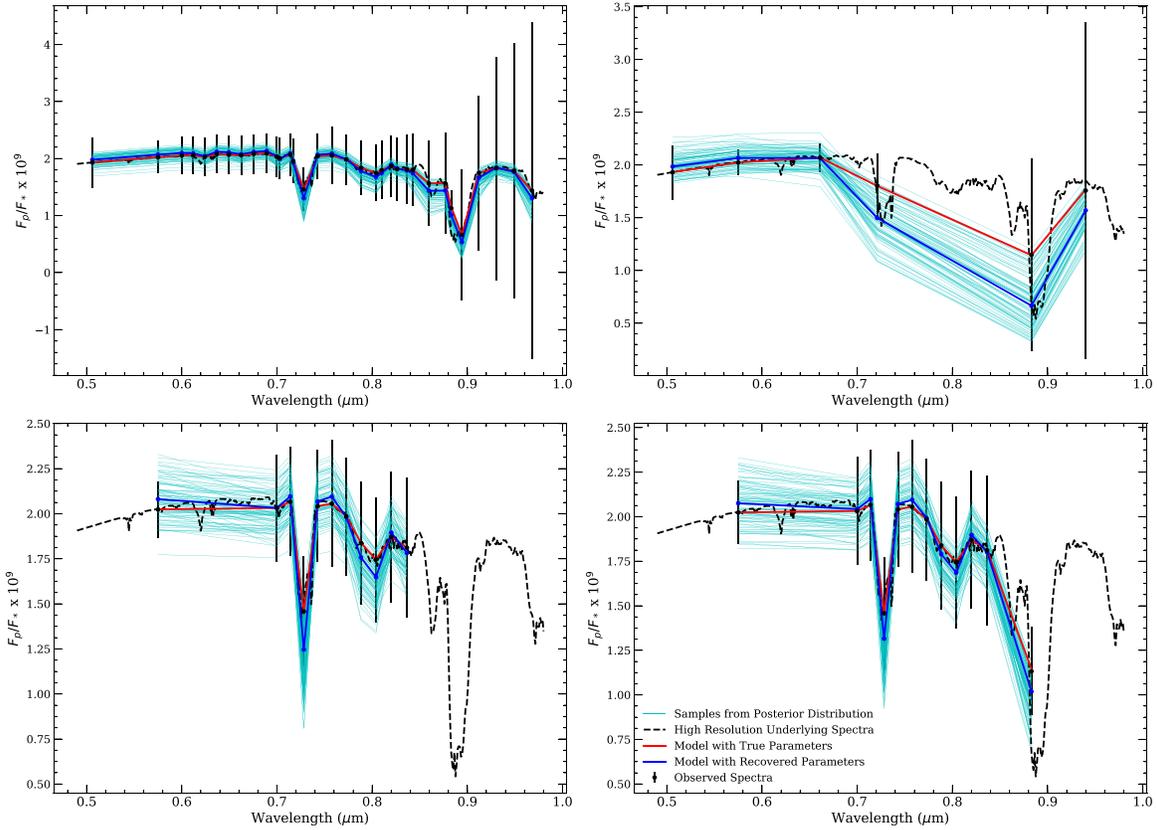

**Figure 15.** Example planet–star flux ratio spectra for our fiducial system (defined at the beginning of Section 5) and for four observing cases (top left-hand panel: case one, top right-hand panel: case two, bottom left-hand panel: case three, bottom right-hand panel: case four—all of the cases are defined in Section 5.2). All spectra are computed assuming a solar-metallicity spectrum from the *CoolTLusty* metallicity grid model class. The total exposure time is 5000 hr for observing cases one, three and four. For observing case two, the total exposure time is 500 hr. The black-dashed line shows the high-resolution underlying spectra. The black dots show the clean signal that would be seen by *WFIRST*-CGI for each observing case in the absence of noise. Black error bars show the noise. Note that where dots do not lie along the high-resolution spectra, they represent an imaging band. The red line connects the true data points to guide the eye. The blue dots and solid line show the model given by the median parameters of posterior distribution mapped by MCMC fitting process. The cyan lines are a random sampling of 100 lines from the MCMC chain. The long-wavelength planet–star flux ratio spectra appears unrealistically well constrained considering the huge error bars. This happens because the metallicity is fitted as a single unit. The shorter wavelength coverage with smaller error bars is sufficient to constrain $R_p$ and metallicity.

### 5.3. Hybrid Model Parameter Recovery

Figure 16 shows some representative posterior distributions for the hybrid model (Section 4.1) parameters as constrained by the four observing cases described in Section 5.2. To demonstrate the data's ability to distinguish the two most extreme types of planet this model can represent, we superpose the posterior distributions of the fully Jovian parameterization with $P_c = 1.0$, $P_m = 1.0$ and the fully Neptunian parameterization with $P_c = 0.0$, $P_m = 0.0$. The underlying planet radius for both cases is 1.0 $R_J$, as stipulated for our fiducial system. Different colors show the posterior distributions for observations done with two different exposure times: 1000 hr and 5000 hr for observing cases one, three and four, 500 and 1000 hr for observing case two. Figure 17 shows the $1\sigma$ error computed from the posterior distribution of each parameter as a function of exposure time, comparing all four observing cases. Due to the form of our model for S/N, we would expect errors in Figure 17 to scale approximately with the inverse square root of the exposure time and eventually asymptote to a minimum value if the parameter in question is well constrained.

Recalling that the shorter wavelength parameter, $P_c$, is meant to be a proxy for chromophore-like absorption, it is not surprising that observing cases three and four do not constrain $P_c$ at all. The posterior distributions in Figure 16 span the whole range of possible values, and the standard deviation of the distribution hovers around 0.3 regardless of exposure time (Figure 17, left-hand panel). With neither the 0.506 $\mu$m band, nor the IFS coverage centered at 0.66 $\mu$m the wavelength coverage is not sensitive to the presence or absence of a chromophore. For observing cases one and two, there is an initial flatter portion of the error versus exposure time curve where large uncertainties hinder the slope between the two imaging bands at 0.506 and 0.575 $\mu$m from constraining $P_c$ well. After 800–1000 hr, the error improves with exposure time as expected. The optimal wavelength range to constrain $P_c$ clearly falls at wavelengths shorter than the *WFIRST*-CGI wavelength coverage where chromophore absorption is a more dominant effect, but, based on the simulations of our fiducial system, the 0.506 $\mu$m band 0.575 $\mu$m band and 0.66 $\mu$m IFS band are somewhat sensitive to the presence or absence of a chromophore if more than one is included.

The results from the hybrid model fits show that *WFIRST*-CGI wavelength coverage and resolution are very sensitive to the degree of methane-like absorption via the proxy parameter $P_m$ and the radius of the planet $R_p$. Error versus exposure times curves for both parameters indicate that the fits are not prior dominated after attaining sufficient S/N, and the parameters are both recovered with acceptable accuracy in exposure times that are deemed to be realistic. Observing case two (narrow band imaging only) performs best until well-past desirable observing times, when observing case one eventually surpasses it. Case three, which uses many less bandpasses but for longer times each, performs very similarly to observing case one for constraining $P_m$ and $R_p$. The flattening of the error versus exposure time curves in Figure 17 reflect the fact that observing case two is approaching the maximum S/N values around 1000 hr, while the other observing cases are approaching their maximum S/N closer to 10,000 hr.

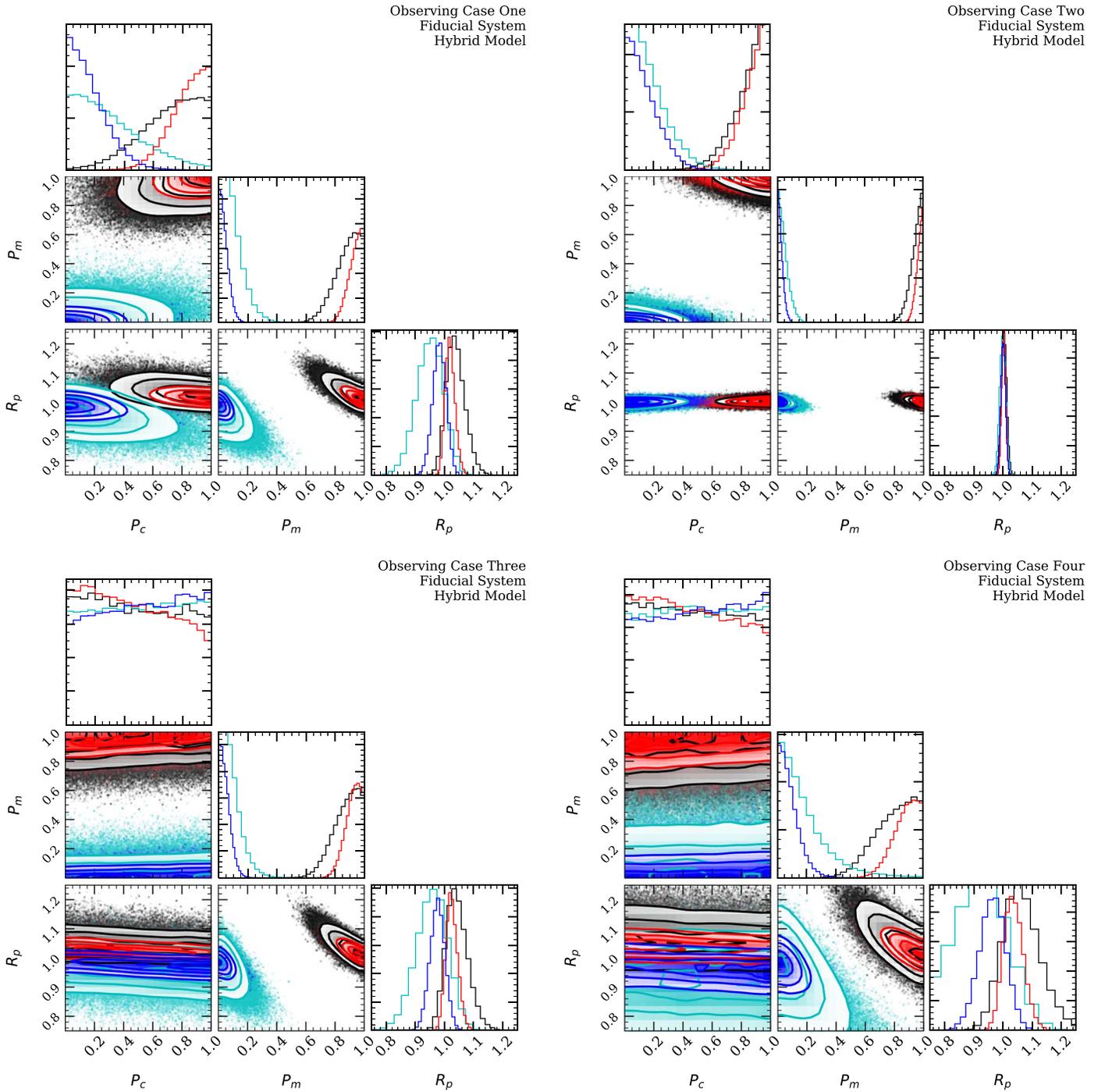

**Figure 16.** Triangle plots showing the posterior distributions of the hybrid model's parameters (chromophore absorption: $P_c$, methane-like absorption: $P_m$, and planet radius: $R_p$, in terms of Jupiter radii) for all observing cases as defined in Section 5.2. The fully Jovian-like parameterization, with $P_c = 1.00$, $P_m = 1.00$, is shown in red (5000 hr exposure/1000 hr exposure for cases one, three and four/case two respectively), and black (1000 hr exposure/500 hr exposure). The fully Neptune-like parameterization, with $P_c = 0.00$, $P_m = 0.00$, is shown in blue (5000 hr exposure/1000 hr exposure), and in cyan (1000 hr exposure/500 hr exposure). Contours mark the $0.5\sigma$, $1\sigma$, $2\sigma$ and $2.5\sigma$ regions, with points outside these regions represented as dots. The histograms at the top of each column show the 1d distribution of the parameter labeled on the $x$-axis below. $P_c$ is not constrained well in any of the observing cases but is particularly bad for cases three and four, which do not include the 0.506 $\mu$m. $R_p$ and $P_m$ show an apparent metallicity-radius degeneracy in all four observing cases but they are both still reasonably constrained to a sub-region of what the priors allow, presumably due to continuum regions. Sufficient S/N and higher spectral resolution can further lessen the degeneracy. These plots and subsequent triangle plots made use of the `corner` Python package (Foreman-Mackey 2016).

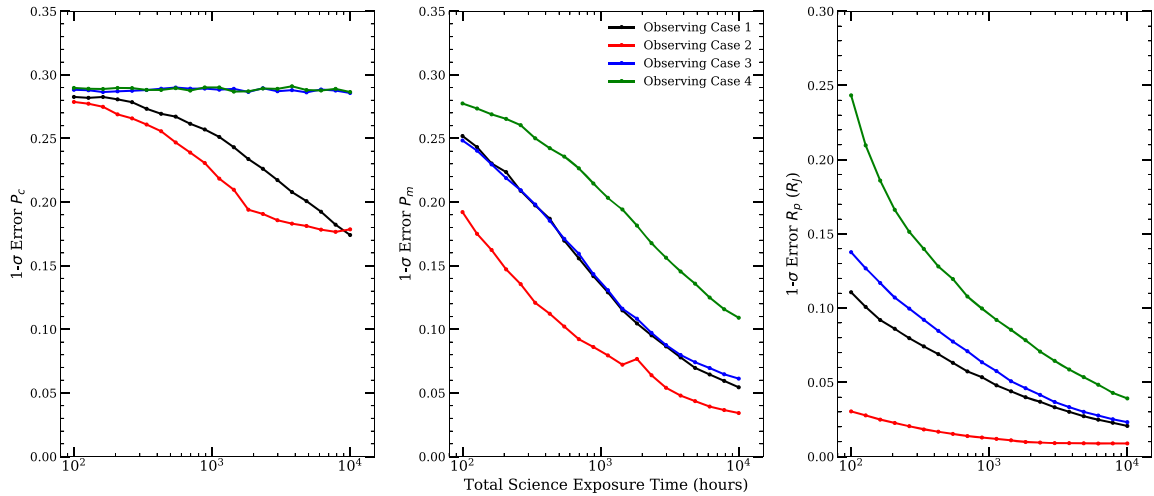

**Figure 17.** 68% confidence interval, or $1\sigma$ error, of the posterior distributions for each hybrid model parameter mapped by our MCMC methods, as a function of total on-target exposure time. The colors of the points and lines denote the four observing cases defined in Section 5.2. For this comparison, we fix planet–star exposure time characteristics to our fiducial system (defined at the start of Section 5) and adopt the parameterization $P_c = 0.5$, $P_m = 0.5$ to minimize the chance of the posterior distributions taking the form of an upper or lower bound rather than a Gaussian distribution about the best fit. A similar study over a grid of $P_c$–$P_m$ parameterizations found that the precise formulation of the geometric albedo spectra was secondary to the exposure time in determining the trends of $1\sigma$ error values.

With our fitting method, the hybrid model suffers from the effects of an albedo-radius degeneracy, tending to overestimate and underestimate $R_p$ to balance out overestimates and underestimates of absorption. Recall that higher $P_c$ values correspond to a more Jupiter-like short-wavelength behavior, meaning more chromophore absorption and a lower albedo. Lower values of $P_m$ correspond to more Neptune-like long-wavelength trends and, thus, more methane absorption and a lower albedo. We expect lower $P_m$ and higher $P_c$ values to correspond to overestimates of $R_p$. By examining the panels in Figure 16, which show $P_m$ and $R_p$ (center of bottom row in each case), it is clear that where $P_m$ is too low, $R_p$ is too high, and visa versa. This degeneracy is less clear between $R_p$ and $P_c$, presumably because $P_c$ has a much smaller effect at wavelengths covered by *WFIRST*-CGI. The $P_m$ and $R_p$ degeneracy does not break completely within exposure times realistic for *WFIRST*-CGI. Despite this degeneracy, $R_p$ is limited to a relatively narrow range of values by the posterior distributions in all three observing cases and is generally slightly better constrained than the other parameters. This happens because the planet–star flux ratio (see Equation (1)) has a steeper dependence on $R_p$ than on the geometric albedo which $P_c$ and $P_m$ set, and the continuum regions of the spectrum are free of the degeneracy. It is important to note that our assumption that the phase function, phase angle, and planet–star separation are all perfectly known certainly lowers the error relative to what a true measurement of planet radius, marginalizing over uncertainties in these other parameters, might have.

Figure 18 shows the standard deviation of the posterior distribution of each parameter as a function of phase angle, assuming both a Lambertian phase function and Mayorga et al.'s (2016) empirical Jovian phase function, comparing the results for all three observing cases. The hybrid model's parameters were fixed to $P_c = 0.75$ and $P_m = 0.25$, and total exposure time was fixed to 1500 hr. As expected, the errors are larger for phases with less illumination, where S/N is lower for a given exposure time. The steep improvement seen for phase angles less than 75° in observing cases one and two is the effect of the 0.506 $\mu$m imaging band coming into a working angle with reasonable coronagraph performance. It is still clear that observing as close to full illumination as possible while maintaining good wavelength coverage can provide significantly better results. Because the Lambertian phase function has a higher planet–star flux ratio at phases between roughly 40° and 110°, assuming an underlying Lambertian phase function when generating and fitting the simulated spectra attains lower errors than the Jovian phase function (compare the dashed and dotted lines).

### 5.4. CoolTLusty: Metallicity and Radius Recovery

By fitting the simulated observations with the *CoolTLusty* metallicity grid, we find that trends in metallicity are recoverable but only with moderate precision at the longest feasible exposure times (and given our strong assumptions that the atmosphere's haze properties are similar to Jupiter, which allows geometric albedo to be set purely by metallicity). Figure 19 shows representative posterior distributions for observing cases one, two, three and four. Once again, we superpose the two most extreme planets that can be represented by the *CoolTLusty* metallicity grid geometric albedo model (solar metallicity exactly and thirty times solar metallicity) observed for 1000 and 5000 hr in observing cases one, three, and four, and for 500 and 1000 hr in observing case two. The data is again generated with an underlying planet radius of 1.0 $R_J$. The radius is then fitted simultaneously with the metallicity.

Looking at Figure 20, which shows the one-sigma error in all four observing cases as a function of exposure time, it is clear that the precision of the metallicity estimate does not scale with the inverse square root of the exposure time until after around 700 hr for observing cases one, three, and four and after around 200 hr for observing case one. At the shorter exposure times, the posterior distribution runs into the priors that metallicity must lie between 1 and 30 times solar making these errors appear smaller than they ought to be. A radius-metallicity degeneracy is apparent in the sloped elliptical shape of the posterior distributions in Figure 19. Despite the partial degeneracy, longer exposure times narrow the allowed range of metallicity, which improves the precision of the radius estimate. Recall that metallicity is directly setting the geometric albedo spectrum in this model. Again, we find that planet radius can be recovered well—within ±15% for over ~1100 hr exposure times. Precision on the $R_p$ estimate scales with the exposure time in the manner expected for our noise model. Observing case three constrains both parameters the best, while observing cases one, three and four all perform similarly well. As in the hybrid model case, our assumption that the phase function, phase angle, and planet–star separation are all perfectly known lowers the error relative to what a true measurement of planet radius, marginalizing over uncertainties in these other parameters, might have.

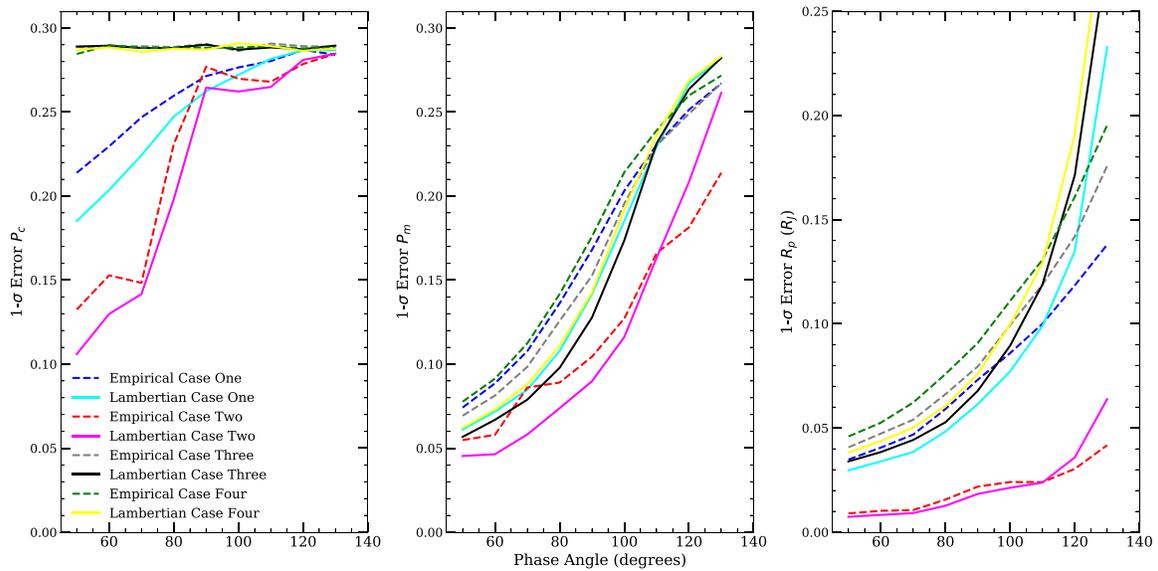

**Figure 18.** 68% confidence interval, or 1σ error, of the posterior distributions for each hybrid model parameter mapped by our MCMC methods, as a function of phase angle. We explore phases ranging from $\alpha = 40°$ to $\alpha = 130°$, and compare the results for a Lambertian phase function (solid) and the empirical Jovian phase function reported in Mayorga et al. (2016) (dashed). The blue and cyan lines represent observing case one, the red and magenta lines represent observing case two, and the black and green lines represent observing case three as defined in Section 5.2. For this comparison, we again fix planet–star system characteristics to our fiducial ~Jupiter–Sun system and adopt the parameterization $P_c = 0.5$, $P_m = 0.5$. We assume a exposure times just below the point that S/N begins to improve less significantly with additional exposure time. These are 100 hr for observing cases one and three and 24 hr for observing case two. The extremely steep jump in errors around $\alpha = 70°–80°$ is the transition where the 0.506 μm band moves from completely within the functioning working angles of the coronagraph through a region of poor performance and eventually completely exterior to the outer working angle. The dip that occurs past 90° for $P_m$ and $R_p$ is also likely due to IFS and imaging coverage at shorter wavelengths moving toward better working angles for coronagraph performance before the effect of significantly less illuminated area overpowers this improvement.

By fixing metallicity at five times solar, we again explore the precision of the *CoolTLusty* metallicity grid parameter recoveries at phase angles ranging from 50° to 130° and compare the results for Lambertian and empirical Jovian phase functions (Figure 21). Our findings are similar to the results from fitting the hybrid model at different phases: errors shrink by over a factor of two between $\alpha = 100°$ and $\alpha = 50°$, and Lambertian phase functions predict lower errors since they overestimate planet flux. We also conducted the same test for all 20 exposure times and we found that errors scale in much the same way with phase at all exposure times, with only the normalization shifting up or down.

### 5.5. Applications to Known RV Planets

Similar to Figures 12 and 11, Figures 22 and 23 depict the results of calculations of S/N as a function of exposure time for the central wavelength of the IFS filters and for the imaging filters, but now using planet radii, planet–star separations, stellar types, stellar magnitudes, and observing distances that correspond to known RV planets (see Table 2 in the Appendix for a list of planet–star system parameters and an explanation of their selection). Our fiducial system is also plotted for comparison. We assume that all planets are observed at phase $\alpha = 65°$ and we use Mayorga et al.'s (2016) empirical Jovian phase function. Planets do not appear in plots where they fall outside the working angles of the appropriate coronagraph for either imaging (the HLC) or spectroscopy (the SPC). Out of these known RV targets, 47 UMa c attains an S/N above 5 in most bands, and is observable across all imaging and IFS bands. Upsilon And d attains higher S/N but may not be visible in the longer wavelength IFS bands. This planet is on an eccentric orbit, and, depending on its argument of pericenter, will have a different apparent separation at $\alpha = 65°$. As Figures 22 and 23 show, a number of other known planets attain S/N similar to the fiducial system for which the previous tests (Sections 5.1, 5.3, and 5.4) were carried out, in at least a subset of the *WFIRST*-CGI wavelength coverage.

To quickly evaluate the potential for observing and characterizing 47 UMa c and Upsilon And d, synthetic observations were generated for all four observing cases at 60° and 90°, at 1000 and 5000 hr for cases one, three and four, and at 500 and 1000 hr for case two, and at the extreme ends of our planet model parameterizations: a Neptune-like planet with $P_c = 0.0$, $P_m = 0.0$, and a Jupiter-like planet with $P_c = 1.0$, $P_m = 1.0$, a solar-metallicity planet, and a 30 times solar-metallicity planet. For 47 UMa c the radius was taken to be just above Jupiter's at 1.15 $R_J$. For Upsilon And d the radius was taken to be 1.0 $R_J$. The posterior distributions for 47 UMa c observed at $\alpha = 60°$ and Upsilon And d at $\alpha = 90°$ are shown in the Appendix in Figures 24 and 25, respectively. Overall, the findings from the fiducial case seem to hold: $R_p$ is constrained and rough trends with the degree of methane absorption measured via $P_m$ or metallicity depending on the model are recovered. $P_c$ is constrained by observing cases one and two, but not by cases three and four.

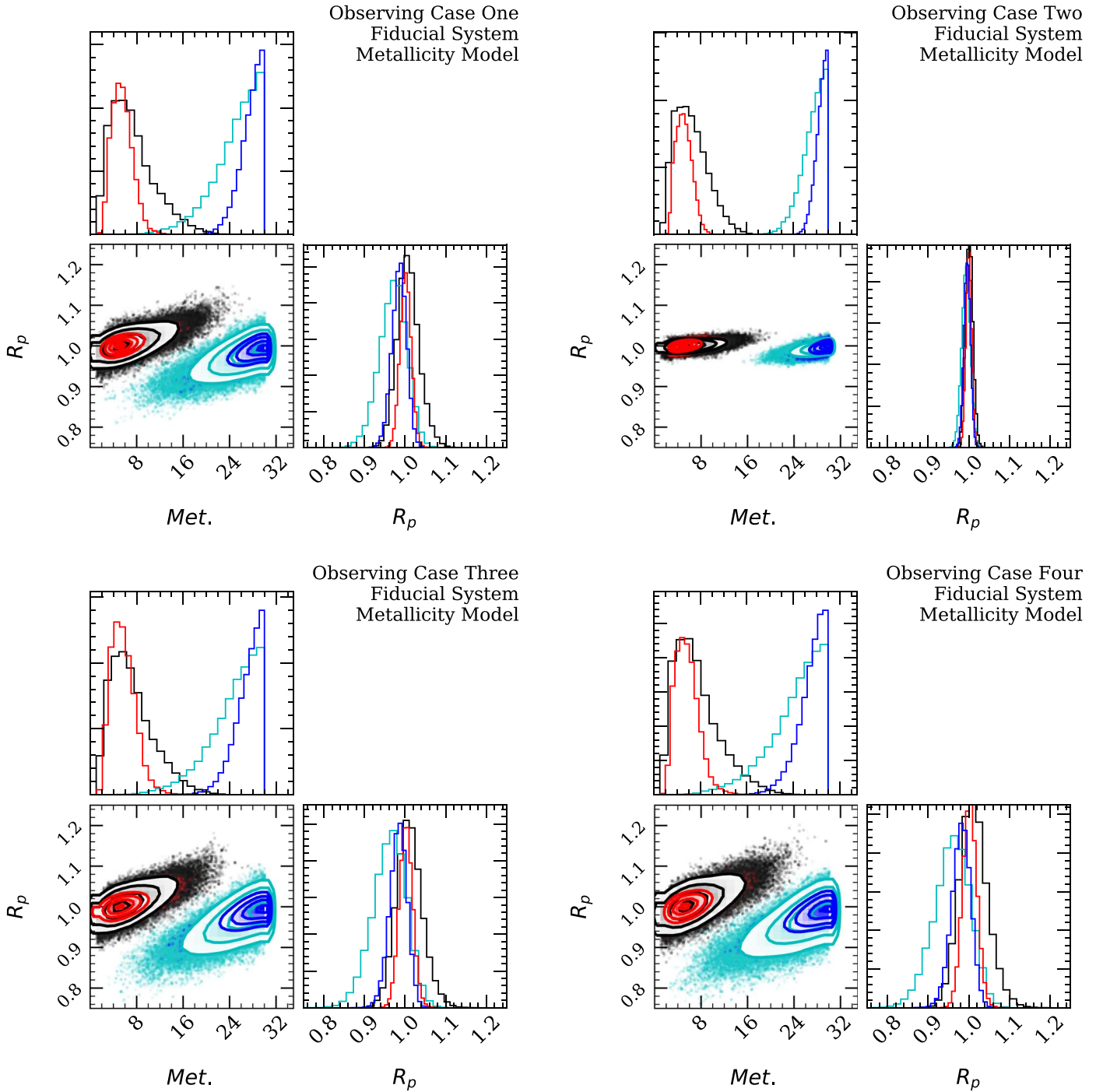

**Figure 19.** Triangle plots showing the MCMC sampling of the posterior distributions of the *CoolTLusty* model parameters (metallicity in terms of solar metallicity and planet radius, $R_p$, in terms of Jupiter radii) for the four observing cases as defined in Section 5.2. The synthetic observed spectra assumed our fiducial system (defined at the start of Section 5). A solar-metallicity parameterization is shown in red (5000 hr exposure/1000 hr exposure for cases one three, and four/case two respectively) and black (1000 hr exposure/5000 hr exposure). A 30-times solar-metallicity parameterization is shown in blue (5000 hr total exposure/1000 hr total exposure) and in cyan (1000 hr exposure/500 hr exposure). Contours mark the $0.5\sigma$, $1\sigma$, $2\sigma$ and $2.5\sigma$ regions, with points outside these regions represented as dots. The histograms at the top of each column show the 1d distribution of the parameter labeled on the *x*-axis below.

In light of these findings for Upsilon And d and 47 UMa c, and for several other systems that achieve S/N levels comparable to, or superior to, our fiducial system seen in Figures 22 and 23, it is sensible to take the previous explorations of the hybrid model and *CoolTLusty* metallicity grid as an assessment of *WFIRST*-CGI performance, provided that the engineers' current best estimates are not over optimistic, and final adopted wavelength coverage is similar to one of the four observing cases defined in Section 5.2.

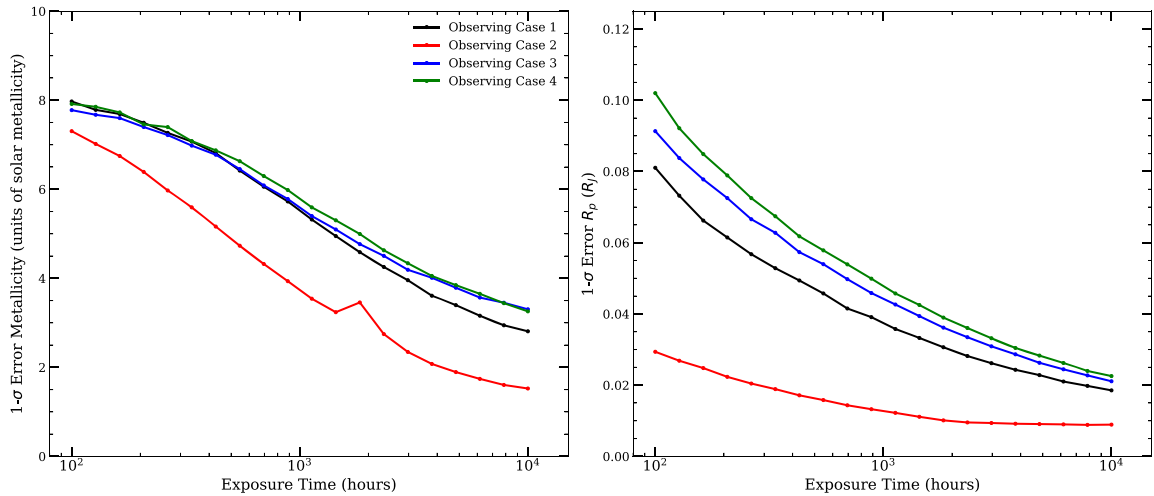

**Figure 20.** 68% confidence interval, or $1\sigma$ error, of the posterior distributions for both *CoolTLusty* model parameters mapped by our MCMC methods as a function of total on-target exposure times ranging from 100 to 10,000 hr. All simulated observations and fits assumed a 15-times solar-metallicity parameterization of the model and our fiducial system's planet–star characteristics. Again, the four observing cases (as defined in Section 5.2) are differentiated by the colors indicated in the left-hand panel's legend. The error on metallicity from observing case two falls below 5 times solar metallicity within 600 hr, the total exposure time for which all individual bandpass observations will fall below the 100 hr limit. For context, Jupiter's metallicity is 3.16 times solar while Neptune's metallicity is closer to 30 times solar. These extremes should be readily distinguishable. For observing cases one, three, and four, the errors on metallicity initially decrease slowly, before steepening around a total exposure time of roughly 700 hr. This is likely due to the priors that metallicity must be between $1\times$ solar and $30\times$ solar limiting the extent of the posterior distributions, which cause smaller standard deviations than the low S/N data at short exposure times can actually constrain on its own.

### 5.6. The Effect of Phase-curve Wavelength Dependence

Previous studies have pointed out that gas giant planets exhibit a change in color with phase. In our previous explorations, we simplified by assuming a wavelength-independent phase function across the *WFIRST*-CGI coverage. To test the impact of this common simplification, we simulated *WFIRST*-CGI observations for the four observing cases assuming the fiducial system, a planet geometric albedo spectrum with $P_c = 0.5$, and $P_m = 0.5$, and a wavelength-dependent Rayleigh-scattering phase curve generated using the methods described in Section 4.1. For observing cases one, three, and four, we incorporated noise assuming a 1000 hr total exposure time. For observing case two, we assumed a total exposure time of 500 hr. We then fitted the four cases in the same manner as described in Section 5.2, assuming a fixed value of Mayorga et al.'s (2016) empirical phase function evaluated at $\alpha = 65°$ regardless of wavelength.

In all cases, the planet radius was recovered consistently with the true value (within the errors). The trends of error with exposure time were similar to the tests carried out previously for all three parameters.

This simple test hints that, for the quality of data expected with *WFIRST*-CGI, our ignorance of the wavelength dependence and precise shape of planet phase curves may not be the dominant problem. This is especially true if exoplanets observed with *WFIRST*-CGI are similar to Jupiter in their scattering properties. The shaded blue region in Figure 7 demonstrates the range of values that a Rayleigh-scattering phase curve can take, depending on the scattering albedo. Around $\alpha = 60°$, the Rayleigh-scattering phase function can range from ~0.42 to ~0.5, depending on the scattering albedo. The fixed value of the empirical phase curve is near the upper end of the Rayleigh range but the Lambertian phase curve is quite a lot higher. It should be noted that real planets are seen to have more extreme color variations than the analytic Rayleigh-scattering phase curves can explain.

## 6. Discussion

### 6.1. WFIRST-*CGI* Exposure Times

The precise combination of imaging and IFS filter coverage that will fly with *WFIRST*-CGI has been an ongoing topic of discussion. An earlier iteration of the mission proposed six imaging bands and three IFS bands represented by observing cases one and two, but the technology demonstration version of *WFIRST* will only fully commission one IFS band centered at 0.76 μm and two imaging bands centered at 0.575 and 0.825 μm, approximately represented by observing cases three and four. Our results relating S/N to exposure times provide support for the mission's decision to shift from the originally proposed three IFS filters to the two IFS filters at the shorter wavelengths, and to change from the narrow 0.883 and 0.940 μm imaging bands to one wider band centered at 0.825 μm. If we want to identify the presence of a chromophore, then fully commissioning the 0.506 μm imaging band and the IFS band centered at 0.66 μm would also be valuable.

As mentioned within Section 5.1, the model that we have implemented for *WFIRST*-CGI noise would change if we consider the case that observations are post-processed with a matched filtering algorithm (Kasdin et al. 2003) rather than reference differential imaging. This could provide improvement over the S/N-exposure time relations presented here. Other instrument and engineering advancements, such as alternative ways of removing cosmic rays, may also influence instrument throughput and favorably improve S/N. The possibility of pairing *WFIRST* with an external occulter (or starshade) would enable a much higher throughput, which also motivates the project to include longer wavelength filters—even if they are not paired with the coronagraph.

The long exposure times will limit the total number of planets that can be characterized by the mission. This will make it difficult to conclusively identify trends with planet size, stellar host type, planet–star separation and metallicity, but it may still be possible to uncover hints and rule out some theories depending on what is observed.

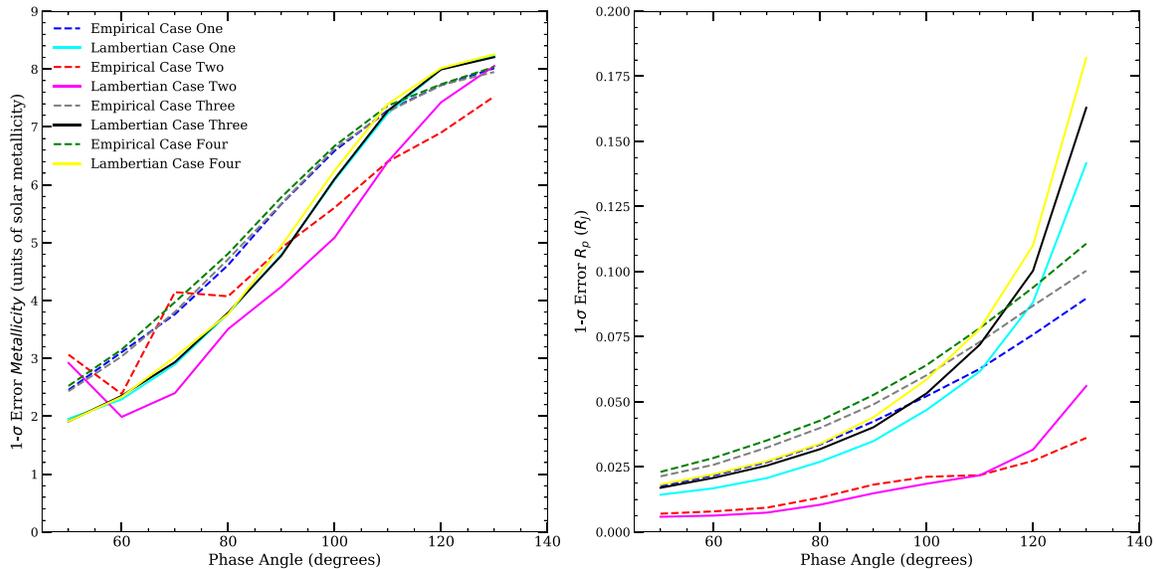

**Figure 21.** 68% confidence interval, or 1σ errors, for the *CoolTLusty* metallicity geometric albedo planet model at phase angles ranging from 50° to 130°. The dashed lines indicate the results for an empirical Jovian phase function and the solid lines indicate the results for a Lambertian phase function. Colors correspond to the observing cases defined in Section 5.2. We fix the total exposure time for each observation to 1500 hr, which means that in observing case one the exposure times for each individual filter are near the maximum desirable exposure times. As planets move from 0° toward 180°, less and less of their illuminated disk is visible to the observer, causing the signal to drop and the errors to increase. The overall trends with phase do not seem to depend heavily on the precise phase function. However, the Lambertian phase function consistently recovers parameters with smaller errors for phase angles below ~110° because it predicts larger phase function values at these phase angles and thus higher S/N. A similar study was carried out for total exposure times ranging from 100 to 10,000 hr, which found that the trends with phase function do not depend heavily on exposure time for their overall shape.

### 6.2. Applicability of our two Geometric Albedo Models

The hybrid model addresses the question of whether or not *WFIRST*-CGI will be able to distinguish a Jupiter-like atmosphere from a Neptune-like atmosphere. More specifically, will *WFIRST*-CGI be able to recognize chromophore-like absorption at short wavelengths or methane-like absorption at long wavelengths? Our findings indicate that the chromophore absorption will be recognizable only if it is near the levels seen on Jupiter and the target system allows S/N around 10 for imaging bands and around 4–5 for IFS bands at shorter wavelengths. Differences in longer wavelength methane-like absorption on the order of that between Jupiter and Neptune will be distinguishable for many of the known-RV planets most likely to be selected as targets.

The *CoolTLusty* metallicity grid addresses the following question: assuming known haze/cloud properties, can we measure the metallicity of *WFIRST*-CGI targets? Our findings indicate that with high enough S/N (S/N ~ 5.0 at IFS 0.66 and 0.76 μm band centers), qualitative trends in metallicity should be recognizable. Even without the IFS, the imaging-only observing case can also distinguish between solar and 30 times solar metallicities. Because planets can have higher metallicities than 30× solar, it would be worthwhile to expand the metallicity grid upwards.

Based upon our fiducial system, 47 UMa c, and Upsilon And d, both the hybrid Jupiter–Neptune model and the *CoolTLusty* metallicity grid would be useful for fitting actual *WFIRST*-CGI data and are of the appropriate complexity for the likely data quality. Given our assumptions, planet radius is well-recovered within ±15% for all observing cases with both the hybrid model and the *CoolTLusty* metallicity grid. Recall that these assumptions included two pieces. First, we assume a well-understood phase function and well-known Keplerian orbital parameters of the target planet, ignoring these sources of uncertainty altogether rather than marginalizing over them. Second, we assume that the geometric albedo spectrum can simply be parameterized, either because cloud/haze properties are similar to Jupiter (the *CoolTLusty* model) or because atmosphere properties lie between the two extremes of Jupiter and Neptune (the hybrid model). Fit results for both models demonstrate that a rough indication of the metallicity based upon the 0.65–0.90 μm planet–star flux ratios should be possible for *WFIRST*-CGI targets, especially with observing case two. Even fitting our two extremely simplified models to synthetic data, which fully meets all our assumptions, realistic S/N levels and wavelength coverage only constrain the other parameters within fractional errors of roughly ±20% ($P_m$ and metallicity) and ±40% ($P_c$, for observing cases one and two) at the $2.5\sigma$ level. In light of this, it seems likely that real observations will be able to confidently differentiate between extreme cases using these models but will lack the precision necessary to uncover subtle trends.

### 7. Summary and Conclusions

In this paper we examined the potential of *WFIRST*-CGI observations to provide information about extrasolar giant planets. To do this, we constructed two simple models for giant exoplanet geometric albedo spectra and paired these with the empirical Jovian phase curve to compute likely observed *WFIRST*-CGI planet–star flux ratio spectra. We then used Nemati et al.'s (2017) parameterized semi-analytic noise model to add noise for four different observing scenarios combining subsets of the imaging filters and IFS coverage. We presented calculations of the S/N for given exposure times for known RV planets. We also demonstrated the general dependence on

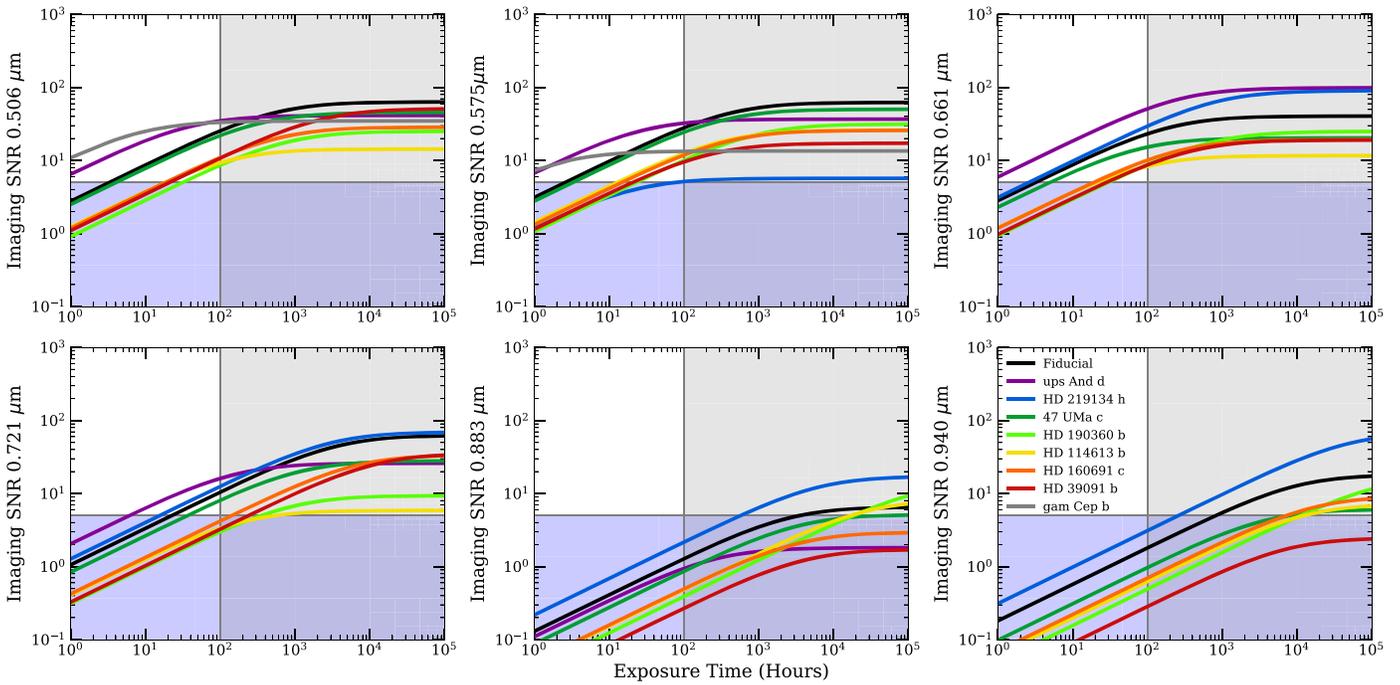

**Figure 22.** S/N dependence on exposure time computed for known RV planets in the originally planned imaging bands (labeled on the *y* axis of each panel). All S/N calculations assume a Jovian geometric albedo spectrum, edge-on orbit, and a phase angle of 65°. We incorporate the eccentricities of the planets, assuming an argument of pericenter in the plane of the sky. We include our adopted fiducial system in black (system defined at the start of Section 5). The gray shaded region indicates exposure times above 100 hr. The purple shaded region indicates S/N below five. Bands centered at 0.506, 0.575, 0.661, and 0.721 μm attain useful S/N for several systems within 100 hr. The longest wavelength imaging bands centered at 0.883 and 0.94 μm do not yield S/N above five for any system in under 100 hr.

relevant star–planet system parameters, including separation, planet radius, distance from observer, phase angle, and exozodi level. We carried out a suite of MCMC fits on simulated data to evaluate the precision and accuracy with which input geometric albedo model parameters and planet radius could be recovered. We also compared our four observing scenarios. We evaluated the effect of assuming a simple wavelength-independent phase function when fitting data generated with a wavelength-dependent phase function for the same four observing cases. Our important findings are summarized in the following list:

1. The imaging bands can attain S/Ns of 15–40 within 100 hr for the 0.506, 0.575, 0.661, and 0.721 μm bands for planet–star system characteristics consistent with known RV planets. The 0.883 and 0.940 μm bands only attain an S/N of 5 or greater for the most favorable cases with large planet radii, small separations, and a nearby planet–star system.

2. Even for promising known RV planets (e.g., 47 UMa c and Upsilon And d), reasonable exposure times for the IFS will be several hundreds of hours, which limits the number of targets observed (likely to only one or two). This is consistent with previous work (Traub et al. 2016).

3. Observing systems that are nearby improves S/N through multiple avenues. In addition to increasing the observed photon flux (which scales with the inverse of distance squared), this allows the target planet to be observed at a smaller separation or closer to full phase without falling interior to the IWA.

4. Observing at a favorable phase angle can lead to ∼factor of two improvements in S/N for both the IFS and imaging, depending on the exact wavelength and bandwidth. This translates into the one-sigma errors on recovered parameters shrinking by a factor of two to four.

5. The main challenge facing the IFS is at the longer wavelengths. Some of the most interesting features occur at these wavelengths but a variety of factors conspire to give this portion of the spectrum an order of magnitude lower S/N for any given exposure time. Many targets move interior to the IWA at those wavelengths, rendering them completely unobservable. Or, if they are observable, targets have lower S/N due to the deep and wide spectral features, plunging quantum efficiency, and increasing noise contributions from the growing PSF-size.

6. Pairing photometry in *WFIRST*-CGI imaging filters with IFS coverage will significantly improve constraining power, particularly at the shortest wavelengths where the IFS does not function and at the longest wavelengths where the IFS struggles to achieve sufficient S/N.

7. Observing planets with radii more similar to Neptune than Jupiter will likely be limited to the nearest stars (within less than 8 pc). Even then, if they exhibit significantly lower geometric albedos than those considered here, they will remain out of sight.

8. Obtaining an S/N of 5 in the IFS at the bottom of the 0.88 μm methane feature will not be possible except for separations and radii which are implausible, based upon known RV targets. This could change if *WFIRST*'s IFS were later paired with a starshade, allowing significantly higher throughput of planet light.

9. The level of exozodi in a system does not have a strong impact on S/N when observing Jupiter-sized planets at 3.8 au, even up to 200 exozodis.

10. For planets similar to the known RV list presented in Table 2, constraining both planet radius and the level of methane absorption is feasible if phase functions and other orbital parameters are known.

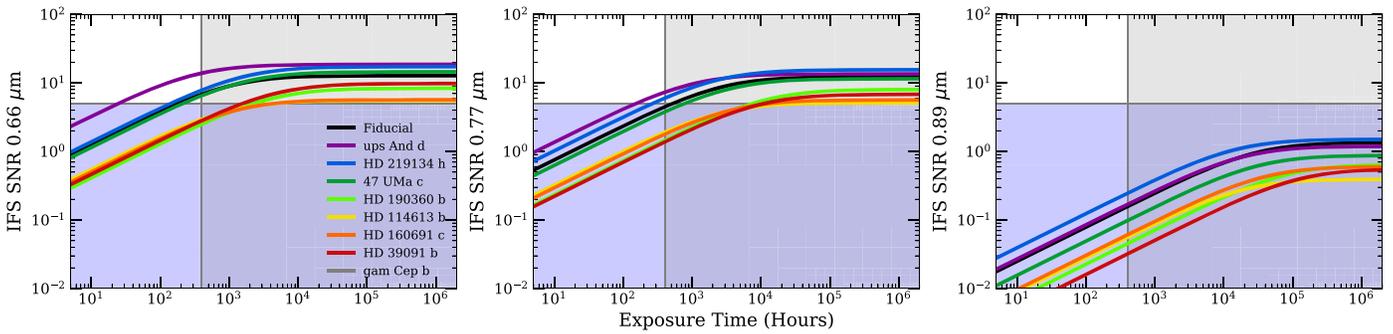

**Figure 23.** S/N dependence on exposure time computed for known RV planets and our adopted fiducial system (black) at the center of the three IFS bands (left: 0.66 $\mu$m, center: 0.76 $\mu$m, right: 0.89 $\mu$m). For simplicity, our calculations assume that all planets have a Jovian geometric albedo spectrum, edge-on circular orbits, and are observed when the planets are at a phase angle of 65°. We incorporate the eccentricities of the planets, assuming an argument of pericenter in the plane of the sky. The gray shaded region indicates exposure times above 400 hr. The purple shaded region indicates S/N below five. These S/N calculations demonstrate the motivation behind leaving off the IFS coverage at 0.89 $\mu$m. Unless targets closer by than the known RV planets are discovered, the longest wavelength coverage of the IFU will not attain reasonable S/N at $R = 50$.

11. Our hybrid Jupiter–Neptune model places constraints on the methane-like absorption parameter $P_m$ with around ±20% accuracy at the 2.5$\sigma$ level. This is sufficient to differentiate between Jupiter and a Neptune-like planet with Jupiter's radius.

12. The wavelength coverage of *WFIRST*-CGI IFS is not optimal for constraining the degree of chromophore-like absorption parameterized with $P_c$, but this information can be recovered from imaging observations if S/Ns around 5–10 can be attained in the short-wavelength imaging bands at 0.506 and 0.575 $\mu$m.

13. Our *CoolTLusty* metallicity grid indicates that trends in metallicity are recoverable assuming that other atmosphere properties are fixed, but only with moderate precision of ±5 times solar metallicity at the one-sigma level and ±10 times solar metallicity at the 2$\sigma$ level. The exception is observing case two, which incorporates all six imaging bands, which is able to achieve significantly higher S/Ns and corresponding higher precisions.

14. Using both models for geometric albedo, we find that planet radius can be recovered within less than ±0.1 $R_J$ and under 15% uncertainty at the 2.5$\sigma$ level. We note that this precision will degrade somewhat when uncertainties in phase angle and phase function are fully accounted for.

15. A Lambertian phase function and the observed Jovian phase function predict systematic differences on a level that is smaller than the uncertainties in the recovered parameters. In cases of the highest S/Ns, it may be necessary to marginalize over the uncertainty in the character of the phase function.

16. Ignoring the wavelength dependence of a Rayleigh-scattering atmosphere's phase curve does not dominate the uncertainties in fits to spectra of the typical quality expected for *WFIRST*-CGI.

It is clear that the *WFIRST*-CGI mission pushes the boundaries of our current technology, making it a vital step forward in space-based high-contrast imaging of exoplanets. While data seem likely to be of limited quantity and only modest S/N, the prospects of obtaining new insights into the characteristics of the few giant extrasolar planets that will be observed in detail seem promising.

The authors would like to acknowledge support for this research under their NASA *WFIRST*-SIT award # NNG16PJ24C, NASA grant NNX15AE19G, and NASA JPL subcontracts no. 1538907, 1529729, 1513640 and 1534432-B-4.25. We are grateful to Bijan Nemati and Jeremy Kasdin for insightful conversations, and we thank Sergei Yurchenko and Alex Howe for help with the ExoMol methane database.

# Appendix
# Noise Model Details

## A.1. Quantum Efficiency, Excess Noise Factor, and Other Losses in the EMCCD

We adopt a wavelength-dependent detector quantum efficiency (QE), which falls off in sensitivity toward redder wavelengths in approximate accordance with Nemati (2014) and Harding et al. (2016):

$$\eta_{\det}(\lambda) = \begin{cases} 0.87, & \lambda < 0.7 \ \mu m \\ 0.87(1.0 - (\lambda - 0.7)/0.32)) & \lambda > 0.7 \ \mu m. \end{cases}$$
(15)

We then multiply $\eta_{\det}$ by additional losses due to hot pixels (0.98), losses due to cosmic rays (0.9), and a net charge transfer efficiency (0.865); all values taken from Nemati et al. (2017). We assume the IFS and imager will both operate in photon-counting mode, so we also multiply by photon-counting efficiency (0.9). The effective value of $\eta$ is thus reduced from ~90% to ~60% at wavelengths shorter than 0.7 $\mu$m. All photonic terms, including zodiacal and planet light, will be reduced by this full value of $\eta$.

The EMCCD can also operate in analog gain mode, not photon-counting mode. The stochastic nature of the amplification process in the EMCCD results in an excess noise factor, or *ENF*, which multiplies contributions to $r_n$ from planet light, speckles, zodiacal light, dark current, and clock-induced change by an amount proportional to the actual rate of the signal. The *ENF* asymptotes to a maximum value $\sqrt{2}$ (Denvir & Conroy 2003). When operating in photon-counting mode, this excess noise factor does not apply.

## A.2. Electronic Effects: Dark Current, Read Noise, and Clock-induced Charge

Our model includes noise from electronic sources originating in the detector. The dark current rate into the signal region is

**Table 2**
Selection of RV Planets Taken from the Confirmed Planet List Curated by CalTech[a] as of 2018 May

| Planet Name | Semimajor Axis (au) | Eccentricity | Planet $m \sin i$ | Distance to system (pc) | Primary Apparent $V$-band Magnitude | Modeled Stellar Type |
|---|---|---|---|---|---|---|
| Fiducial System | 3.80 | 0.0 | 1.0 | 10.0 | 5.0 | G0 V |
| Upsilon And d | 2.50 | 0.3 | 4.132 | 13.47 | 4.1 | F5 V |
| HD 219134 h | 3.11 | 0.06 | 0.34 | 6.55 | 5.57 | K5 V |
| 47 UMa c | 3.60 | 0.1 | 0.54 | 14.08 | 5.05 | G0 V |
| HD 190360 b | 3.97 | 0.31 | 1.54 | 15.89 | 5.73 | G5 V |
| HD 114613 b | 5.16 | 0.25 | 0.48 | 20.48 | 4.85 | G0 V |
| HD 160691c | 5.24 | 0.1 | 1.814 | 15.28 | 5.15 | G0 V |
| HD 39091 b | 3.38 | 0.64 | 10.27 | 18.21 | 5.65 | G0 V |
| gam Cep b | 2.05 | 0.05 | 1.85 | 13.79 | 3.21 | K0 V |

**Notes.** This table includes only those systems that are shown in Figures 22 and 23. They are loosely chosen to be those targets likely to attain adequate S/N with *WFIRST*-CGI, i.e. systems closer to Earth, and to fall within the working angles of the coronagraphs at most wavelengths. There are other confirmed planets that would attain similar or slightly lower S/N values in the full wavelength coverage and planets that would attain higher S/N values but only in a subset of the wavelength coverage. The list was truncated for visual clarity in Figures 22 and 23.
[a] https://exoplanetarchive.ipac.caltech.edu

given by:

$$r_{\text{dark}} = i_d N_{\text{pix}}^{\text{detector}}, \qquad (16)$$

where $i_d$ is the level of dark current present in the detector with units of $e^-/s/\text{pixel}$. Over the course of the mission, the levels of dark current are expected to rise as radiation damage degrades the detector (Harding et al. 2016). Preliminary tests predict a factor of two increase in dark current levels from the beginning of mission to the end of the mission. For our purposes, we adopt a level of dark current that is consistent with the telescope having spent three months at L2.

For an IFS, we calculate the number of pixels involved in the signal region, $N_{\text{pix,detector}}$, according to:

$$N_{\text{pix}}^{\text{detector}} = N_{\text{spec}} \left( N_{\text{lens}} \frac{\lambda}{\lambda_c} \right)^2, \qquad (17)$$

where $N_{\text{spec}}$ is the number of pixels in each $\Delta\lambda$ in the spectral dimension, $N_{\text{lens}}$ is the number of lenslets that the signal falls on in one spatial dimension, and $\lambda_c$ is the wavelength at which the system has been designed to be Nyquist sampled. The IFS will be designed to have constant dispersal, making $N_{\text{spec}}$ independent of wavelength. Taking our fiducial values for $N_{\text{spec}}$ and $N_{\text{lens}}$ we have:

$$N_{\text{pix}}^{\text{detector}} \sim 45 \left( \frac{\lambda}{\lambda_c} \right)^2. \qquad (18)$$

Imaging signal regions contain ~4 times less pixels because there is no spectral dispersal, scaling instead as:

$$N_{\text{pix}}^{\text{detector}} \sim 11 \left( \frac{\lambda}{\lambda_c} \right)^2. \qquad (19)$$

These pixel number calculations are certainly not a perfect representation of what the *WFIRST*-CGI mission will ultimately do. In practice some form of matched filtering algorithm will likely be used to extract the signal. We assume the use of aperture photometry in such a way that we best match the results obtained by more detailed end-to-end modeling (Rizzo et al. 2017). The imaging pixel scale will be 0.021 arcsec/pixel, and the integral field spectroscopy pixel scale will be 0.029 arcsec/pixel.

Clock-induced charge, which is associated with high-frequency clocking needed for rapid readout rates, is accounted for as:

$$r_{\text{cic}} = q_{\text{cic}} \frac{N_{\text{pix}}^{\text{detector}}}{t_{\text{fr}}}, \qquad (20)$$

where $q_{\text{cic}}$ is the average amount of clock-induced charge in units of $e^-/\text{frame}/\text{pixel}$, and $t_{\text{fr}}$ is the maximum length for a single exposure in the observing sequence in units of seconds. Generally, $t_{\text{fr}}$ is chosen to limit the probability of cosmic-ray hits to the detector during an exposure.

Levels of both dark current and clock-induced charge will be reduced by the losses from hot pixels, cosmic rays, charge transfer efficiency and, in the case of the IFS, photon-counting efficiency, similar to the photonic sources.

The read noise is described by:

$$r_{\text{read}} = i_r \frac{N_{\text{pix}}^{\text{detector}}}{t_{\text{fr}}}, \qquad (21)$$

where $i_r$ is the effective read noise in units of $e^-/\text{frame}/\text{pixel}$. $i_r$ is really the more conventional understanding of CCD read noise $\sigma_{e^-}$, attenuated by the electron multiplying register's gain stages, $G_{\text{EM}}$. For the detector baselined for *WFIRST*'s coronagraph $\sigma_{e^-}$ is on the order of tens of $e^-/\text{frame}/\text{pixel}$ and $G_{\text{EM}}$ is on the order thousands, giving an effective read noise much less than one (Harding et al. 2016).

In the case of photon-counting mode, levels of read noise will be additionally reduced by the photon-counting efficiency but not by the cosmic rays, hot pixels, or charge transfer efficiency. This happens because it enters into the data after all of the gain stages.

### A.3. Throughputs

We must specify three separate throughputs because a coronagraph affects planet light ($\tau_{\text{pla}}$), zodiacal light ($\tau_{\text{zod}}$), and speckle light ($\tau_{\text{spe}}$) differently. Point sources, extended sources, and residual scattered light all travel different optical paths within the instrument. Obscuration due to the struts/secondary mirror, the losses from reflections, losses due to the filter, and losses due to the polarizer affect all three throughputs and they are accounted for in our collecting area: $A_{\text{PM}} = \tau_{\text{obs}}(\pi/4)D^2$, where $\tau_{obs}$ is the throughput accounting for obscuration from struts and secondary mirror. The planet throughput must also account for the occulting mask and PSF core throughput, combined into the term $\tau_{core} = \frac{\tau_{occ}}{\tau_{PSF}}$. This gives the expression for planet throughput:

$$\tau_{pla} = \tau_{core}\, \tau_{ref}\, \tau_{fil}\, \tau_{pol}, \qquad (22)$$

where $\tau_{ref}$ is the throughput accounting for losses from reflections, $\tau_{fil}$ is the throughput accounting for losses due to the filter, and $\tau_{pol}$ is the throughput accounting for losses due to a polarizer. Note that the core throughput and occulting throughput will depend on the working angle at which the planet falls onto the detector. To incorporate this working angle dependence into our model, we employ numerical coronagraph performance models from Krist et al. (2016), as shown in Figure 1. When computing rates for the IFS, we use the SPC table produced in 2017 June. When computing rates for the imager, we use the HLC table created in 2016 December.

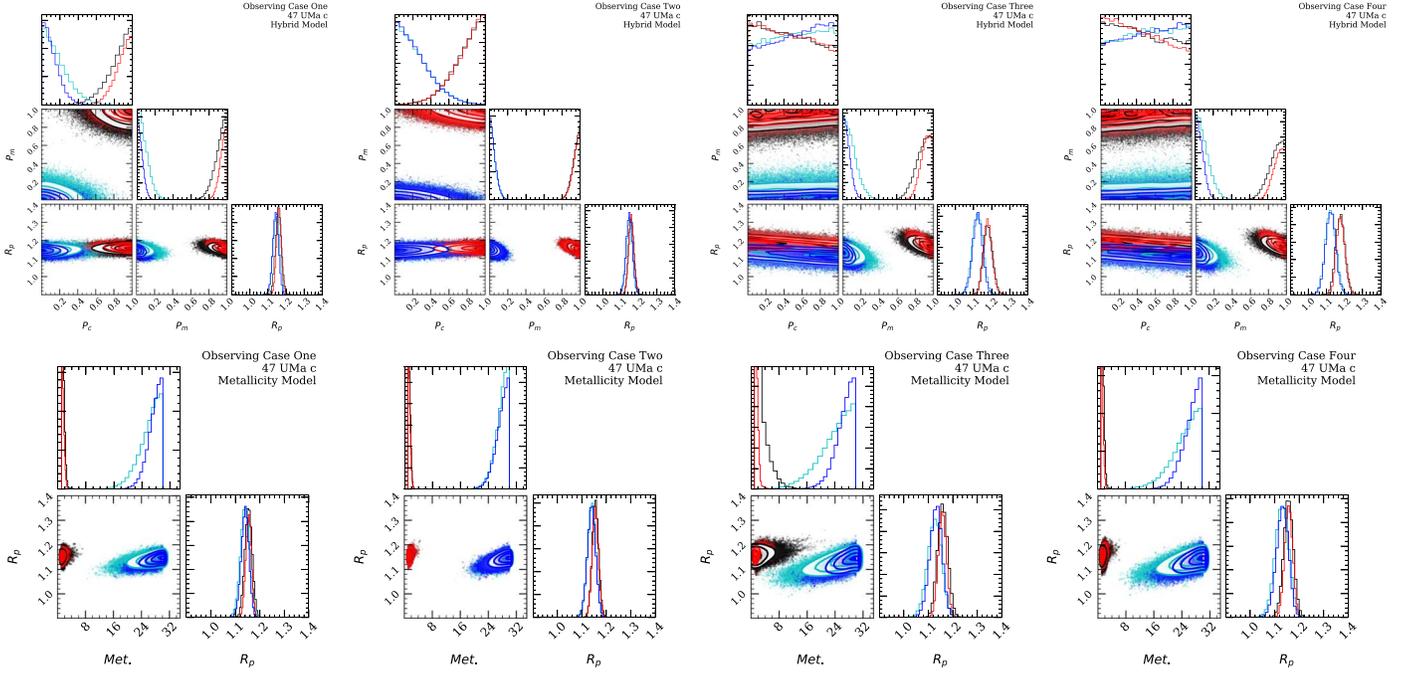

**Figure 24.** Triangle plots showing the MCMC sampling of the posterior distributions of the hybrid model's parameters (top row) and the *CoolTLusty* model parameters (bottom row) for the four observing cases (cases one through four run left to right, observing cases are defined in Section 5.2), carried out for a planet–star system with the inferred characteristics of 47 UMa c listed in Table 2 and a planet radius of 1.15 $R_J$. The top row is similar to Figure 16, with the fully Jovian-like parameterization, $P_c = 1.00$, $P_m = 1.00$, shown in red (5000 hr total exposure/1000 hr total exposure for cases one, three and four/case two respectively) and black (1000 hr total exposure/500 hr total exposure). The fully Neptune-like parameterization, $P_c = 0.00$, $P_m = 0.00$, is shown in blue (5000 hr total exposure/1000 hr total exposure) and in cyan (1000 hr total exposure/500 hr exposure). The bottom row is similar to Figure 19, with a solar-metallicity parameterization shown in red and black, and a 30-times solar-metallicity parameterization is shown in blue and in cyan. All these figures have the same exposure times as the hybrid model. Contours mark the 0.5σ, 1σ, 2σ and 2.5σ regions, with points outside these regions represented as dots. Like our fiducial system, it is assumed that 47 UMa c is observed at a phase angle of 60° and exhibits the empirical Jovian phase curve reported by Mayorga et al. (2016). The fits are more precisely inferred for 47 UMa c than for our fiducial system, particularly the *CoolTLusty* model's metallicity parameter.

The zodiacal light is a diffuse source, so it need only account for the occulting mask without considering core throughput:

$$\tau_{zod} = \tau_{occ}\, \tau_{ref}\, \tau_{fil}\, \tau_{pol}. \qquad (23)$$

The reason for this difference in diffuse sources is that any light scattered *out* of the signal area due to core throughput limitations will be made up for by light scattered *into* the signal area from another part of the sky. This throughput thus assumes that the zodiacal light is of uniform brightness across the field of view. Although this assumption is likely to be wrong, there is little alternative because the exact structures of the dust disks are not known.

The speckle throughput leaves off losses due to the occulter because those are accounted for within the coronagraph performance model itself, and we do not want to double count them. It also leaves off losses due to core throughput because speckles arise from scattered light within the optics. It is, thus, defined as:

$$\tau_{spe} = \tau_{ref}\, \tau_{fil}\, \tau_{pol}. \qquad (24)$$

### A.4. Planet Count Rate

The planet count rate is the product of the collecting area $A_{PM}$, planet throughput $\tau_{pl}$, quantum efficiency $\eta$, and the incoming flux density of photons from the planet. This flux density is simply the planet–star flux ratio given by Equation (1) multiplied by the stellar flux. To model stellar flux we scale spectra of various main-sequence stellar types at zero magnitude to the desired absolute *V*-band magnitude, and then account for observing distance. This is similar to the methods employed by Nemati et al. (2017). For the IFS, the calculation is as follows:

$$F'_{*,\lambda} = F_{0,\lambda} \left(\frac{10\,\text{pc}}{d_{obs}}\right)^2 10^{-M_*/2.5} \frac{\lambda^2}{hcR}, \qquad (25)$$

where $F_{0,\lambda}$ is the standard zero magnitude V-band flux of the desired spectral type, $d_{obs}$ is the distance to the observed system, $M_*$ is the absolute magnitude of the host star in *V* band, $\lambda$ is the wavelength of light, *h* is Planck's constant, and *c* is the speed of light. We take advantage of the constant resolving power $R = \frac{\lambda}{\Delta\lambda}$ of the IFS to convert the specific flux density to a photon flux. To compute rates for the imager, we carry out a similar calculation to the IFS but now integrate across the spectral range that falls within the imaging band.

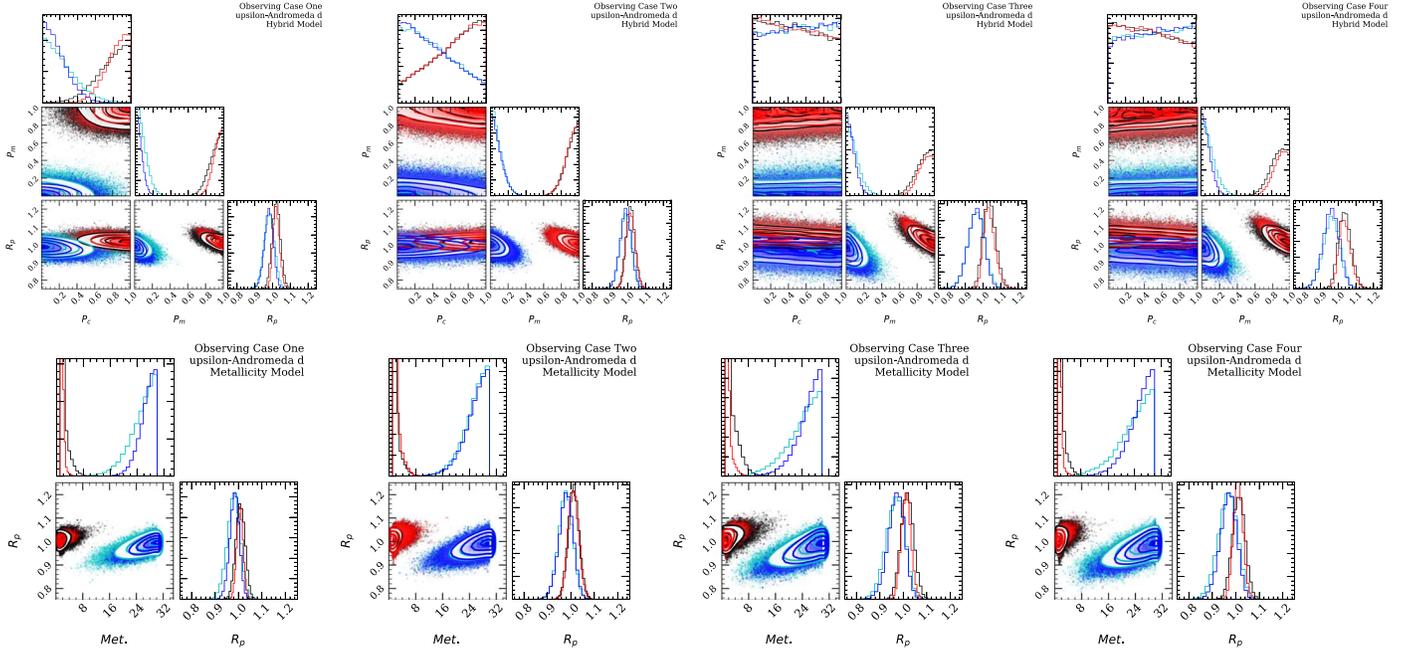

**Figure 25.** Triangle plots showing the MCMC sampling of the posterior distributions of the hybrid model's parameters (top row) and the *CoolTLusty* model parameters (bottom row) for the four observing cases (cases one through four from left to right, as labeled), all carried out for a planet–star system with the inferred characteristics of Upsilon And d listed in Table 2, and a planet radius of 1 $R_J$. Model parameterizations and exposure times are the same as Figures 24, 16, and 19. It is assumed that Upsilon And d is observed at a phase angle of 90° to minimize the wavelength coverage which falls interior to the inner working angle of the coronagraph. Upsilon And d has a semimajor axis of only 2.52, and is nearer by than 47 UMa c, but it also has a large eccentricity. If the argument of pericenter is favorably oriented, then it may actually be observable at more highly illuminated phase angles than 90°. In all cases, recovery of the Upsilon And d model parameters has similar results to the fiducial system and 47 UMa c.

### A.5. Zodiacal and Exozodiacal Count Rate

We include background light from zodiacal and exozodiacal dust following Robinson et al.'s (2016) adaptation of Stark et al. (2014). In particular, Stark et al. (2014) demonstrated that a constant *V*-band surface brightness of $M_{z,V} = 23$ mag arcsec$^{-2}$ provides a reasonable representation of zodiacal light ignoring the true variation with ecliptic latitude and longitude first presented by Levasseur-Regourd & Dumont (1980). The count rate from zodiacal light is:

$$r_z = A_{PM}\tau_{zodi}\eta\Omega\frac{F_{\odot,\lambda}(1\,\text{au})}{F_{\odot,V}(1\,\text{au})}F_{0,V}10^{-M_{z,V}/2.5}\frac{\lambda^2}{hcR}, \quad (26)$$

where $\Omega$ is the wavelength-dependent angular size of the photometry aperture area in arcsec$^2$, $M_{z,V}$ is the magnitude of zodi surface brightness, $F_{\odot,\lambda}(1\,\text{au})$ is the specific flux density of the Sun at a distance of 1 au, $F_{\odot,V}(1\,\text{au})$ is the flux density of the Sun at 1 au in *V* band, and $F_{0,V}$ is the flux density of a zero magnitude star in *V* band.

Stark et al. (2014) define a "zodi" as the surface brightness of an exozodiacal disk a distance of 1 au from a solar twin. This has exozodi surface brightness $M_{ez,V} = 22$ mag arcsec$^{-2}$, approximately double the local zodi because the observer receives reflected light from above and below the midplane of the system. The count rate from a system with $N_{ez}$ zodis of dust, for a planet *a* au from its star is given by:

$$r_{ez} = A_{PM}\tau_{zodi}\eta\Omega N_{ez}\left(\frac{1\,\text{au}}{a}\right)^2 F_{\star,\lambda}'(1\,\text{au})\frac{F_{0,V}10^{-M_{ez,V}/2.5}}{F_{\odot,V}(1\,\text{au})}, \quad (27)$$

where $F_{\star,\lambda}'(1\,\text{au})$ is the photon flux density a distance of 1 au from the star (e.g., put 1 au into Equation (19) for $d_{obs}$). Since $\Omega$ will scale with $\lambda^2$ both $r_z$ and $r_{ez}$ are proportional to $\lambda^4$.

### A.6. Speckle Count Rate

Following the work of Krist et al. (2016) and Nemati et al. (2017), we adopt the speckle count rate:

$$r_{sp} = A_{PM}\tau_{sp}\eta F_{\star,\lambda}' C_{raw} I_{pk} N_{pix}^{model}. \quad (28)$$

The coronagraph raw contrast, $C_{raw}$, is the azimuthally averaged, smooth background contrast that is achieved at a given radius on the detector. The normalized peak pixel value, $I_{pk}$, is the throughput into the peak pixel of the psf from any given point in the field. Both values are are taken from tables produced by Krist et al. (2016), the same models are shown in Figure 1. We multiply $I_{pk}$ by $N_{pix}^{model}$, while conservatively assuming the worst case scenario for the speckle light—that is, every pixel in the signal area has the peak value. $N_{pix}^{model}$ is a value distinct from the number of pixels in the signal region on the actual detector, $N_{pix}^{detector}$, because the tables are created by assuming a different pixel size than may eventually be used on the real instrument.

## ORCID iDs

B. Lacy  https://orcid.org/0000-0002-9420-4455
A. Burrows  https://orcid.org/0000-0002-3099-5024